\documentclass[useAMS,usenatbib,usegraphicx]{mn2e}
\usepackage{fixltx2e} 
\usepackage{aas_macros}     
\usepackage{times}          
\usepackage[dvips]{color}   
\usepackage[dvipdfm]{hyperref} 

\usepackage{ifthen}         
\usepackage[normalem]{ulem} 

\newcommand{\annotate}{true}   

 \ifthenelse{\equal{\annotate}{true}}{
   \newcommand{\pbtodo}[1]{ [\textbf{\textcolor{red}{#1}}] }
 }{
   \newcommand{\pbtodo}[1]{}
 }

\newcommand{\hrsim}{hMS}              
\newcommand{\lcdm}{$\Lambda$CDM}      %
\newcommand{\vv}[1]{\bmath{#1}}       
\newcommand{\mat}[1]{\mathbfss{#1}}   
\newcommand{\dotprod}{\bmath{\cdot}}
\newcommand{\crossp}{\bmath{\times}}
\newcommand{\rhocrit}{ \rho_\rmn{crit} }
\newcommand{\Deltac}{ \Delta_\rmn{c} }
\newcommand{\Omegatot}{\Omega_\rmn{tot}}
\newcommand{\Omegab}{\Omega_\rmn{b}}
\newcommand{\OmegaM}{\Omega_\rmn{M}}
\newcommand{\OmegaL}{\Omega_\Lambda}
\newcommand{\Msol}{\rmn{M_{\sun}}}   
\newcommand{\Npart}{N_{\rmn{part}} }
\newcommand{\Np}{ N_\rmn{p} }
\newcommand{\Nsel}{ N_\rmn{sel} }
\renewcommand{\mp}{ m_\rmn{p} }
\newcommand{\Rvir}{R_\rmn{vir}}
\newcommand{\Vvir}{V_\rmn{vir}}
\newcommand{\Mvir}{M_\rmn{vir}}
\newcommand{\Lbox}{ L_{\rmn{box}} }
\newcommand{\jp}{j_\rmn{p}}
\newcommand{\jr}{j({\le r)}}
\newcommand{\jin}{j_\rmn{inner}}

\newcommand{\vjin}{\vv{j}_\rmn{inner}}  
\newcommand{\vjtot}{\vv{j}_\rmn{tot}}   
\newcommand{\jsc}{\tilde{\jmath}}       
\newcommand{\vc}{v_\rmn{c}}
\newcommand{\bfrac}{f_\rmn{bary}}
\newcommand{\dd}{\mathrm{d}}            

\newcommand{\Mpc}{\rmn{Mpc}}            
\newcommand{\kpc}{\rmn{kpc}}            
\newcommand{\kms}{\rmn{km\,s^{-1}}}     
\newcommand{\Hunit}{ \, \kms \, \Mpc^{-1}} 
\newcommand{\munit}{ \, h^{-1} \Msol }     
\newcommand{\lunit}{ \, h^{-1} \Mpc }      
\newcommand{\klunit}{\, h^{-1} \kpc }      

\defcitealias{Bett07}{B07}
\defcitealias{BS05}{BS05}

\title[Angular momentum of haloes and galaxies]{The angular momentum
of cold dark matter haloes with and without baryons}

\author[Bett et al.]{Philip Bett,$^{1,2}$\thanks{Email: p.e.bett@dunelm.org.uk}
  Vincent Eke,$^{1}$ Carlos S. Frenk,$^{1}$ Adrian Jenkins,$^{1}$
  \newauthor and Takashi Okamoto$^{1,3}$\\ 
  $^{1}$Institute for Computational Cosmology, University of Durham,
  South Road, Durham, DH1 3LE, UK\\
  $^{2}$Argelander-Institut f\"ur Astronomie, Universit\"at Bonn, 
Auf dem H\"ugel 71, D-53121 Bonn, Germany\\
  $^{3}$Center for Computational Sciences, University of Tsukuba
1-1-1, Tennodai, Tsukuba, Ibaraki 305-8577, Japan
  }

\begin{document}
\date{\today}

\pagerange{\pageref{firstpage}--\pageref{lastpage}} \pubyear{2009}

\maketitle

\label{firstpage}

\begin{abstract}
  We investigate the magnitude and internal alignment of the angular
  momentum of cold dark matter haloes in simulations with and without
  baryons. We analyse the cumulative angular momentum profiles of
  hundreds of thousands of well resolved haloes in the Millennium
  simulation of Springel et al.  and in a smaller, but higher
  resolution, simulation, in total spanning 5 orders of magnitude in
  mass. For haloes of a given mass, the median specific angular
  momentum increases with radius as $\jr\propto r$. The direction of
  the vector varies considerably with radius: the median angle between
  the inner ($\la 0.25\Rvir$) and total ($\le\Rvir$) angular momentum
  vectors is about $25\degr$. To investigate how baryons affect halo
  spin, we use another high-resolution simulation, which includes gas
  cooling, star formation and feedback. This simulation produces a
  sample of galaxies with a realistic distribution of disc-to-total
  ratios, $D/T$: two thirds of the galaxies have $D/T>0.5$ in the
  B-band. The formation of the galaxy spins up the dark matter within
  $0.1\Rvir$ such that the specific halo angular momentum increases by
  $\approx 50$ per cent in the median. The dark matter angular
  momentum becomes better aligned, but there remains a broad
  distribution of (mis-)alignments between the halo and the central
  galaxy, with a median angle between their angular momenta of $\sim
  30\degr$. Galaxies have a range of orientations relative to the
  shape of the halo: half of them have their minor axes misaligned by
  more than $45\degr$, although only about $10$ per cent of the
  galaxies lie within $30\degr$ of the plane perpendicular to the
  major axis of their halo.  Finally, we align a sample of haloes
  according to the orientation of their galaxies and stack the
  projected mass distributions.  Although the individual haloes are
  significantly aspherical, galaxy--halo misalignments produce a
  stacked mass distribution that cannot be distinguished from
  circular. If the lack of alignment found in our simulations is
  realistic, it will be extremely difficult for weak lensing studies
  to measure the ellipticity of cold dark matter haloes using this
  technique.
\end{abstract}



\section{Introduction}
The formation of structures in the universe is often studied in the
context of the two stage model of \cite{whiterees1978}, within the
$\Lambda$ cold dark matter (\lcdm{}) cosmology \citep{whitefrenk91}:
primordial mass density perturbations (dominated by cold dark matter
or CDM) collapse under gravity, hierarchically forming ever larger
haloes, while galaxies form from the baryons that cool and collapse
within them. The first part of the problem -- determining the
structure and evolution of dark matter haloes -- can be tackled
realistically and reliably using $N$-body simulations. Starting from
the early simulation work of \cite{defw85} and
\cite{Frenk85,1988ApJ...327..507F}, there is now a vast amount of
literature on the subject, covering both detailed studies of a few
individual objects at very high resolution (e.g. \citealt*{NFW96},
\citealt{1999ApJ...516..591O}, \citealt{2000MNRAS.317..630K},
\citealt{2001MNRAS.321..559B}, \citealt{2005Natur.435..629S},
\citealt*{2007ApJ...667..859D}, \citealt{2008MNRAS.391.1685S}), as
well as statistical studies of large samples of less well resolved
haloes (e.g. \citealt[(hereafter \citetalias{Bett07})]{Bett07},
\citealt{Hahn2007}, \citealt{Maccio2007}).  Some of these studies have
focused specifically on the angular momentum structure of
haloes, which is the main topic of this paper
(e.g. \citealt{defw85}, \citealt{1987ApJ...319..575B},
\citealt{1992ApJ...399..405W}, \citealt{1996MNRAS.281..716C},
\citealt{2001ApJ...555..240B}, \citealt{2002MNRAS.332..325P,
    2002MNRAS.332..339P}, \citealt{2002MNRAS.336...55C},
\citealt{2002ApJ...581..799V}, \citealt{2005ApJ...629..781K},
\citealt{BS05}, \citealt{2005ApJ...634...51A},
\citealt{2006ApJ...646..815S}, \citealt{2006MNRAS.367.1781A},
\citetalias{Bett07}, \citealt{Maccio2007,Maccio2008}).

The dissipative baryonic processes that produce the visible galaxy are
considerably more complex, and are correspondingly less well
understood, than the purely gravitational processes that make the dark
matter halo. Nevertheless, recent $N$-body/gasdynamic simulations,
which include various forms of feedback between the cooling gas and
the forming galaxy, are beginning to produce fairly (although not
entirely) realistic disc galaxies from \lcdm{} initial conditions.
Recent examples of such work include
\cite{2003ApJ...591..499A,ANSE2003}, \cite*{2003ApJ...596...47S},
\cite{2004ApJ...606...32R},
\cite{2004ApJ...607..688G,2007MNRAS.374.1479G}, \cite{Takashi05},
\cite{2006ApJ...639..126B}, \cite{2006ApJ...641..878T},
\cite*{2006PhRvD..74l3522G}, \cite*{2007ApJ...671..226H},
\cite{2008IAUS..245...33C}, \cite{Croft2009},
\cite{2008ApJ...680.1083R}, \cite{2009IAUS..254..445G},
\cite{2009MNRAS.396..696S}, \cite{2009ApJ...702.1250R}; see also
\cite{2008ASPC..393..111O} and \cite*{2008ASL.....1....7M} for recent
reviews.

Baryonic processes modify the internal structure of the halo in which
they take place, including basic properties such as their angular
momentum and shape profiles. These are the quantities that we study in
this paper. We first exploit the enormous statistical power of the
Millennium Simulation \citep{2005Natur.435..629S} to investigate pure
dark matter haloes with a large range of masses, from galactic to
cluster haloes. We then compare the shape and angular momentum
properties of galactic size haloes with those of haloes in a smaller,
higher resolution simulation that includes baryonic physics. In
addition to the intrinsic importance of these fundamental properties,
we are interested in this problem for two practical reasons. Firstly,
even the best gasdynamics simulations to date appear to suffer from
the ``angular momentum problem'' first highlighted by
\cite{1991ApJ...380..320N}. This is a statement of the fact that
in simulations, excessive transfer of angular momentum from the
cooling gas to the  
halo leads to discs that are much smaller than
observed. On the other hand, calculations of galaxy formation using
semianalytic techniques, which assume that that the baryons initially
have the same angular momentum distribution as the dark matter, and
that this is conserved as the baryons collapse 
produce better agreement with observed disc sizes
\citep[see, e.g.][]{1980MNRAS.193..189F, 1998MNRAS.295..319M,
    Cole2000, 2000ApJ...545..781D, 2007ApJ...654...27D,
    2009MNRAS.397.1254G}.

The second reason why we are interested in halo shapes and angular
momentum is the prospect of measuring the shapes of real galactic
haloes using weak gravitational lensing \citep[see][for a recent
review]{2008ARNPS..58...99H}. In principle, the shapes of dark matter
haloes provide an important test of the \lcdm{} model. Simulations
that neglect baryon effects have demonstrated that cold dark matter
haloes are generically triaxial \citep{1988ApJ...327..507F,
1991ApJ...378..496D, 1992ApJ...399..405W, 2002ApJ...574..538J,
2006MNRAS.367.1781A, 2006ApJ...646..815S, Bett07, Hayashi2007}.  On
the other hand, some alternative gravity theories such as TeVeS/MOND
\citep{Bekenstein2004,2001MNRAS.327..557M} predict the gravitational
potential at large distances from galaxies to be spherical. Since the
gravitational lensing signal from an individual object is very weak,
lensing studies are normally performed on stacked images around large
samples of galaxies, orientated according to the shape of the
galaxy. Intrinsic to this method is the assumption that the shape of
the galaxy and the shape of the halo are tightly correlated with each
other. It is not clear, however, that cold dark matter theory predicts
this.

Recent attempts to measure halo shapes from weak lensing data have been
inconclusive. \cite{HYG2004}, using the Red Sequence Cluster Survey
(RCS), claimed to detect a definite halo ellipticity;
\cite{Mandelbaum2006}, using data from the Sloan Digital Sky Survey
(SDSS), were unable to obtain a definite detection of halo
ellipticity, but their data suggested different galaxy--halo alignment
distributions for spiral and elliptical galaxies.
\cite{Parker2007}, using the CFHT Legacy Survey, 
again detected an aspherical halo shape, but at relatively low
significance.  \cite{Evans2008}, using a similar method but applied to
cluster haloes, claimed to rule out spherical configurations.

There are, of course, other methods for measuring halo shapes,
including using cluster X-ray data in conjunction 
  with the Sunyaev--Zel'dovich effect
  \citep[e.g.][]{2000A&A...364..377R, 2001ApJ...561..600Z,
  2004ApJ...601..599L, 2006ApJ...645..170S}, or examining the
  distribution of satellite galaxies within a halo
  \citep[e.g.][]{1991MNRAS.249..662P, 2000MNRAS.316..779B,
  2008MNRAS.385.1511W}. These are not considered in this paper.

Several studies have investigated the influence of baryonic processes
on the shapes and angular momentum of haloes. The case of
non-radiative gas was considered by \cite{2002ApJ...576...21V} and
\cite{2005ApJ...628...21S}, who found a broad distribution of dark
matter/gas orientations. \Citet{2002ApJ...576...21V} also found
significant misalignment between the inner and outer regions of the
halo.  \cite{Kazantzidis2004}, \cite*{2004IAUS..220..421S}, and
  \cite{Bailin05} all showed that haloes become much more spherical
in simulations with gas cooling and star formation, particularly in
the central regions, but with a significant effect throughout the
halo. This radial trend is the opposite of that found in simulated
haloes without baryons: haloes simulated with just dark matter
become \emph{less} spherical (more prolate) towards their centres
\citep{1991ApJ...378..496D,
    Hayashi2007}. The constrained realisation simulations of
\cite{2006ApJ...648..807B} showed that it is the growth of the
baryonic disc that reduces the halo prolateness, pushing it closer to
spherical\footnote{This behaviour is not universal however. In the
  single object simulated by \cite{2007ApJ...671..226H}, the inner
  regions of the halo are consistently more spherical than the outer
  halo, regardless of whether baryons were included or not.}.  More
recently, \cite{2009arXiv0902.2477A} compared $13$ haloes simulated
with and without gas (including radiative cooling, but without star
formation or feedback), and broadly confirmed these results.

Early investigations of the alignment between the
\emph{angular momentum} vectors of the halo and gas distributions
were performed by
  \cite{2002ApJ...576...21V}, \cite{2003ApJ...592..645Y}, and
  \cite*{2003ApJ...597...35C}, all of whom  considered both cooling and
  non-radiative gas. They found that the
cold gas develops a broader range of orientations relative to the dark
matter than either the hot or non-radiative gas. \cite{Bailin05}
considered the alignment of simulated galaxies with their parent
halo. They found that, although the dark matter at the virial radius
remained essentially uncorrelated with the orientation of the galactic
disc, the presence of the disc had caused the minor axis of the inner
halo to align with the disc axis. \cite*{2006PhRvD..74l3522G}
similarly examined the density, shape and orientation profiles of
haloes in simulations with and without baryonic physics. They too
found that the baryons made their haloes more spherical, with a wide
range of angles between the dark matter and the galaxies.
\Citet{2003MNRAS.346..177V} considered the effect of
  preheating of the intergalactic medium and found that it greatly
  increased the misalignment between the angular momentum of the
  (non-radiative) gas and the dark matter halo.

\cite{Croft2009} used a model that included black hole feedback in the
star formation process, and found a broad distribution of angles
between the orientations of the galaxies and their parent haloes, as
well as between the shape and angular momentum axes of each halo
component (gas, stars and dark matter).  Most recently,
\cite{2009ApJ...702.1250R} compared simulations of a single halo,
with and without baryons and star formation, and found that the dark
matter angular momentum was very well aligned to the stellar component
throughout the halo, but the gaseous component was $\ga100\degr$ out.
However, the alignment of the angular momentum vectors in their halo
varied dramatically as the halo evolved.

While these simulations tend to produce broadly consistent results,
they cannot easily be compared because of the different type of
baryonic processes that they included. An important limitation of
these studies is the small samples of objects they were able
to simulate, which preclude robust statistical statements. The one
exception to this is the large simulation by \cite{Croft2009} which,
however, had to be stopped at $z=1$ due to computing time
restrictions.

In the first part of this paper, we exploit the statistical power of
the Millennium Simulation \citep{2005Natur.435..629S} to determine the
properties of the angular momentum profiles of dark matter haloes of a
large range of mass in a purely cold dark matter universe. In
Section~\ref{s:mshr1}, we consider both the magnitude and orientation
of the angular momentum vectors as functions of halo radius and mass.
In Section~\ref{s:dmodmg}, we include the effects of baryons and
explore a pair of simulations of a smaller volume, one of which
contains just dark matter and the other which contains dissipative
baryons, star formation and feedback as well. We examine the angular
momentum profiles of the haloes in detail, comparing the dark matter
at different radii between the two simulations, and their orientation
relative to the central galaxy.  We also examine the physical reasons
behind the changes induced by the baryons in the dark matter.
Finally, we apply our results directly to the problem of weak
gravitational lensing by computing the 2-D projected mass
distributions (individual and stacked), when haloes are aligned
according to their galaxy's orientation.  We present our conclusions
in Section~\ref{s:conclusions}.


\section{Dark matter haloes}\label{s:mshr1}

We begin by  examining the angular momentum structure of simulated dark
matter haloes without baryons.


\subsection{The simulations}
We analyse two dark matter simulations of the large-scale structure of
a \lcdm{} universe in order to obtain very precise statistics of dark matter
halo properties over a wide range of masses.  The first of these, the
\emph{Millennium Simulation} (MS), has over $10$ billion particles
($2160^3$) in a $500\lunit$ cubic volume; it is described fully in
\cite{2005Natur.435..629S}.  The second simulation (which we will
refer to as \hrsim{}) assumes the same cosmology, but has higher
resolution and a smaller volume\footnote{The \hrsim{} simulation was
also analysed by \cite{neto07} and \cite{2008MNRAS.387..536G}.}.  Both
simulations were carried out with a version of the GADGET-2 code
\citep{gadget2} that was specially optimised for massively
parallel computations and low memory consumption.

We perform all our analyses at redshift $z=0$.  The parameters of the
assumed cosmologies are given in Table~\ref{t:cosparams} and the
simulation parameters in Table~\ref{t:simparams}. (These tables also
give parameters for other simulations used in Section~\ref{s:dmodmg}.)

\begin{table}
  \begin{center}
    \begin{tabular}{l c c c c c c}\hline
      Sims.     & $\OmegaL$ & $\OmegaM$ & $\Omegab$ & $h$    & $n$ &$\sigma_8$\\
      \hline\hline
      MS \& \hrsim{} & $0.75$ & $0.25$  & $0.045$   & $0.73$ & $1.0$ & $0.9$ \\
      DMG \& DMO     & $0.70$ & $0.30$  & $0.044$   & $0.70$ & $1.0$ & $0.9$ \\
      \hline
    \end{tabular}
    \caption{Cosmological parameters (at $z=0$) for the simulations
      used in this paper.  The cosmological density parameters are
      defined as $\Omega_i:= \rho_i / \rhocrit$, where the critical
      density $\rhocrit:= 3H_0^2 / (8\pi G)$, the equivalent mass
      density of the cosmological constant is $\rho_\Lambda:= \Lambda
      c^2/ (8\pi G)$, and $\Omegatot = \OmegaL +\OmegaM =1$.  The
      Hubble constant at $z=0$ is parametrised as $H_0=100h \Hunit$.
      The spectral index is given by $n$, and $\sigma_8$ is the linear
      theory mass variance in spheres of radius $8\lunit$ at $z=0$.
      (The DMO and DMG simulations are described in
      Section~\ref{s:dmodmg}.)}
    \label{t:cosparams}
  \end{center}
\end{table}

\begin{table}
  \begin{center}
    \begin{tabular}{lcrr@{.}lr@{.}l}\hline
      Sim.      & $\Lbox$   & $\Npart$  & \multicolumn{2}{c}{$\mp$} & \multicolumn{2}{c}{$\eta$} \\
                & $\lunit$  &           & \multicolumn{2}{c}{$10^7\munit$} & \multicolumn{2}{c}{$\klunit$}\\
      \hline\hline
      MS        & $500$     & $10\,077\,696\,000$ & 86&07     & 5&0\\
      \hrsim{}  & $100$     & $    729\,000\,000$ &  9&52     & 2&4\\
      \hline
      DMO       & $35.33$   & $      3\,397\,215$ &  1&93     & 0&7\\
      \hline
      DMG:      & $35.33$   & \multicolumn{5}{l}{}                \\
      --DM      &           & $      3\,397\,215$ &  1&65    & 0&7\\
      --Gas     &           & $      2\,985\,242$ &  0&28    & 0&35\\
      --Stars   &           & $      1\,668\,836$ &  0&06    & 0&35\\
      \hline
    \end{tabular}
    \caption{Parameters for the simulations used in this paper: box
      size, numbers and masses of particles, and gravitational
      softening $\eta$ (see equation \ref{e:softkernel}).  For DMG and
      DMO, we analyse the high resolution region in the centre of the
      $\Lbox$ cube; at $z=0$, this region is approximately spherical
      with a diameter of about $12.5\lunit$.  Note also that in DMG,
      the number of gas and star particles, and their masses, vary
      over the course of the simulation according to the star
      formation algorithm.  The values presented here are the numbers
      and median masses at $z=0$.}
    \label{t:simparams}
  \end{center}
\end{table}


\subsection{Halo definition}\label{s:mshr1halodef}
We shall be looking at various halo properties defined in spherical
shells, so it is appropriate to adopt a spherical overdensity
\citep[SO,][]{LC94} algorithm to define the haloes.  In
practice, our halo definition starts by applying a friends-of-friends
algorithm \citep[FOF,][]{defw85}, with a linking length of $0.2$, to
construct an initial set of particle groups.  We then use the
\textsc{Subfind} program \citep{2001MNRAS.328..726S} to identify 
self-bound structures within each group (one of which will always be the
main halo itself), as well as the location of the gravitational
potential minimum.  We identify the potential minimum of the main 
self-bound structure within each FOF group with the centre of the
corresponding dark matter halo.  We then grow a spherical boundary
around each centre until the total enclosed mass density (not just the
original FOF particles) matches that of a virialised halo in the
spherical top hat model for a flat cosmology
\citep*[$\Omegatot=\OmegaM+\OmegaL=1$;
see][]{1996MNRAS.282..263E,1998ApJ...495...80B}:
\begin{equation}
  \Deltac=\frac{\rho}{\rhocrit} \approx 18\pi^2+82(\OmegaM(z)-1) -39(\OmegaM(z)-1)^2.
  \label{e:virdens}
\end{equation}
This gives $\Deltac\approx 94$ for MS and \hrsim{}.  We shall refer to
this halo boundary as its virial radius, $\Rvir$, and to the total
mass within this radius as $\Mvir$.


\subsection{Analysis of physical properties}\label{s:phys}

The kinetic and potential energies of each halo ($T$ and $U$
respectively) are computed as in \citetalias{Bett07}, that is,
\begin{equation}
  T = \frac{1}{2} \sum_{i=1}^{\Np} m_i \vv{v}_i^2, 
\end{equation}
where the halo consists of $\Np$ particles; particle $i$ has mass
$m_i = \mp$, and velocity vector $\vv{v}_i$ relative to the
centre-of-mass velocity.

To calculate the potential energy of each halo, we use a random sample
of up to $N_\rmn{sel}=1000$ particles from each halo, and scale the
total according to
\begin{equation}
  U = \left( \frac{\Np^2-\Np}{\Nsel^2-\Nsel} \right)
      \left( \frac{-G\mp^2}{\eta} \right)
      \sum_{i=1}^{\Nsel-1} \sum_{j=i+1}^{\Nsel} -W_2(r_{ij}/\eta).
      \label{e:pot}
\end{equation}
We take the form of the potential used in the simulation, which
incorporates the SPH smoothing kernel \citep*{2001NewA....6...79S}:
\begin{equation}
  W_2(x) =
     \left\{
        \begin{array}{ll}
           \frac{16}{3}x^2 -\frac{48}{5}x^4 +\frac{32}{5}x^5 -\frac{14}{5}, &
	   0\leq x \leq\frac{1}{2}, \\

           \frac{1}{15x} +\frac{32}{3}x^2 -16x^3 +\frac{48}{5}x^4 \\
	   \;\;\; -\frac{32}{15}x^5 -\frac{16}{5}, &
	   \frac{1}{2}\leq x\leq1, \\

           -\frac{1}{x}, &  x\geq 1. \\
	   
        \end{array}
     \right.
     \label{e:softkernel}
\end{equation}
Here, the argument $x$ is given by the ratio of the particle pair
separation, $r_{ij}$, to the spatial softening length, $\eta$ (see
Table \ref{t:simparams}).  We only use the potential (and kinetic)
energy of a halo to determine whether or not it is in equilibrium, in
order to define the sample of halos to be studied, as explained in the
following subsection. The limits we adopt for this criterion are
relatively broad so the loss of accuracy introduced by random sampling
the halo has no significant impact on the selection.  Indeed, when using
the smaller, higher resolution simulations in Section~\ref{s:dmodmg},
we use \emph{all} particles without random sampling, and the results
agree well with those in this section. 

For computing the specific angular momentum profiles, we divide each
halo into a spherical `inner' core region and a series of concentric
spherical shells, spaced by $0.2$ in
$\log_{10}(r/\Rvir)$, where $r$ is the radial distance from the halo
centre.  We define an inner region  that has
radius $0.1\Rvir$ for the analysis of angular momentum magnitudes, but
radius $10^{-0.6}\Rvir\approx 0.25\Rvir$ for the analysis of
orientations. This larger region is required because the determination
of angles between angular momentum vectors is subject to different
numerical errors from the determination of the magnitude of the
vector, and requires more particles within a given radius to ensure
robust and reliable results (see the discussion in Appendix
\ref{a:cosboot} for full details).

The (cumulative) specific angular momentum vector,
$\vv{j}({\leq r})$, of the $\Np({\leq r})$ dark matter particles within a
given radius $r$ (of total mass $M({\leq r})$) is then given by
\begin{equation}\label{e:jinr}
  \vv{j}(\leq r) = \frac{1}{M(\leq r)}
                  \sum_{i=1}^{\Np(\leq r)} m_i\vv{x}_i\crossp \vv{v}_i,
\end{equation}
where $\vv{x}_i$ and $\vv{v}_i$ are the position and velocity vectors
of particle $i$ relative to the halo centre and centre-of-mass
velocity.  The cumulative specific angular momentum magnitude
  profile is then given by $\jr = |\vv{j}({\leq r})|$. Since we use
the SO algorithm to define our haloes, the total halo specific angular
momentum is $\vjtot = \vv{j}({\leq\Rvir})$.  

Our choice to investigate the cumulative rather than the differential
angular momentum  differs from many previous studies 
\citep[e.g.][]{BS05}.  We made this choice for practical reasons, in
order to maximize the size of our sample of well-resolved haloes: if a
certain minimum number of particles is required for a reliable
measurement of angular momentum, then the use of cumulative profiles
implies that this condition only has to be satisfied in the innermost
bin considered, rather than in each bin separately.  (We discuss our
halo selection criteria in detail in the following subsection).  Of
course, using cumulative profiles has the drawback that the
measurements in each bin are not independent, and bin-to-bin variation
is reduced compared to the differential case.  In practice, however,
the angular momentum is dominated by the mass at large radii, so there
is very little difference between the behaviour of the cumulative
$\jr$ and the differential $j(r)$.  We will show this explicitly in
Section~\ref{s:dmodmg}.

The cumulative angular momentum, $\jr$, defined above also differs
from the quantity considered in \cite{2001ApJ...555..240B}, the
cumulative \emph{mass} profile of specific angular momentum,
$M(<j_z)$.  Here, $j_z$ is the magnitude of the specific angular
momentum projected along the direction of the total angular momentum
vector of the halo.  Any internal misalignment of the haloes will
break the correspondance between this quantity and $\jr$.

The shapes of the dark matter haloes are computed using the inertia
tensor, $\mat{I}$, which directly relates the angular momentum
vector, $\vv{J}$, and the angular velocity
vector, $\vv{\omega}$
(i.e. $\vv{J}=\mat{I}\dotprod\vv{\omega}$).  This tensor has components
\begin{equation}
  I_{\alpha\beta} = \sum^{\Np}_{i=1} m_i \left( \vv{x}_i^2
  \delta_{\alpha\beta} - x_{i,\alpha}x_{i,\beta} \right) 
  \label{e:inertia}
\end{equation}
such that $J_\alpha = I_{\alpha\beta}\omega_\beta$ ($i$ indexes
particles, $\alpha$ and $\beta$ are the tensor indices with values of
1, 2 or 3, and $\delta_{\alpha\beta}$ is the Kronecker delta).  The
eigenvectors of the diagonalised inertia tensor define an ellipsoid,
which represents the equivalent homogeneous shape of the object in
terms of a semimajor axis, $\vv{a}$, intermediate axis, $\vv{b}$ and
semiminor axis, $\vv{c}$.\footnote{As noted in \citetalias{Bett07},
  this gives the same axes as the mass distribution matrix $\mat{M}$
  with components $\mathcal{M}_{\alpha\beta} = \sum^{\Np}_{i=1} m_i
  x_{i,\alpha}x_{i,\beta}$.  The two are related through
  $I_{\alpha\beta}=\rmn{Tr}(\mat{M})\delta_{\alpha\beta} -
  \mathcal{M}_{\alpha\beta}$ \citep[e.g.][]{2008gady.book.....B}.}

We also calculate the angular velocity magnitude profile, $\omega(r) =
|\vv{\omega}(r)|$. The angular velocity at radius $r$ is defined from
the distribution of the mass within that radius through the
expression,
\begin{equation}
  \vv{\omega}(r) = \mat{I}^{-1}(\leq r)\dotprod\vv{J}(\leq r).
  \label{e:angvel}
\end{equation}


\subsection{Halo selection}\label{s:milhalosel}

In their analysis of the halo shapes and spins in the MS,
\citetalias{Bett07} excluded from their halo catalogue haloes that
were clearly out of equilibrium at the time of the simulation
snapshot.  This was achieved by restricting haloes to have an
instantaneous `virial ratio' of energies in the range:

\begin{equation}\label{e:virrat}
  Q := \left| \frac{2T}{U}+1\right| \le Q_\rmn{lim}
\end{equation}
Here, we apply a cut of the same form and adopt the same value of
$Q_\rmn{lim}=0.5$.

\begin{figure}
  \includegraphics[width=80mm]{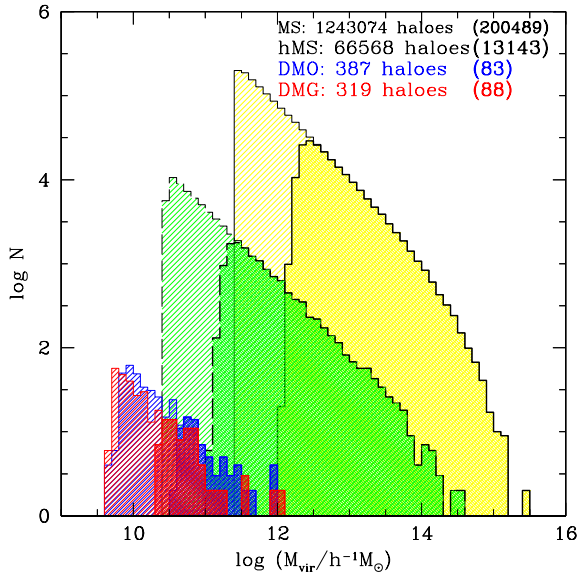} 
  \caption{Mass functions (halo number histograms) for haloes from the
    four simulations used in this paper.  The Millennium Simulation is
    shown in black with yellow shading, \hrsim{} in dashed-black with
    green shading, DMO in blue and DMG in red. (The DMO and DMG
    simulations are described in Section~\ref{s:dmodmg}.)  For each
    simulation, we show the histogram of total mass within $\Rvir$
    (i.e. including stars and gas for the DMG haloes).  We show two
    selections for each simulation: objects that contain at least
    $300$ dark matter particles, within $\Rvir$ (thin lines, light
    shading), and within $0.1\Rvir$ (heavy lines and shading).  In
    both cases, selected haloes must also satisfy the `virialisation'
    criterion, $Q\le 0.5$.}  \label{f:massfns}
\end{figure}

\citetalias{Bett07} found that angular momentum and shape parameters
of a halo were subject to numerical biases if it contained fewer than
approximately $300$ particles.  For the analyses presented here, we
also adopt the $\Np \geq 300$ selection criterion either for the object
as a whole (when profile information is not required), or for the
innermost radial bin considered, e.g. $r<0.1\Rvir$.  This ensures that
only reliable angular momentum profiles are retained and is, by far,
the most stringent criterion applied in this paper.  The mass
functions of haloes selected in this way are shown in Fig.~\ref{f:massfns}.

We apply further selection criteria when investigating the orientation
of the angular momentum vectors since the magnitude of the vector is
related to the uncertainty in the direction.  These are detailed in
Appendix \ref{a:cosboot}. The number of haloes in each sample are
quoted in the legend of the appropriate plots.


\subsection{Results}

\subsubsection{Angular momentum profiles}\label{s:millangmommag}
\begin{figure}
  \includegraphics[width=80mm]{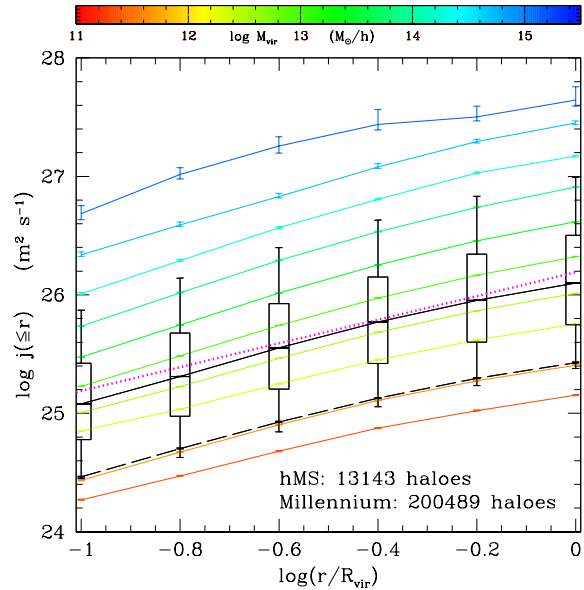} 
  \caption{Dark matter halo cumulative specific angular momentum
    profiles, for objects in the MS and \hrsim{} simulations.
    Coloured lines show the median profiles for haloes in different
    mass bins, and the black lines show the medians for the MS (solid)
    and \hrsim{} (dashed) data together.  The error bars on each line
    (nearly invisible for MS and \hrsim{}) are given by
    Eqn.~\ref{e:mederr}.  The outer bars and boxes on the MS line
    indicate the spread of individual halo profiles around the median
    (the boxes enclose $68$ per cent of the data, the bars enclose
    $95$ per cent).  The dotted magenta line shows the $j\propto r$
    scaling (with arbitrary normalisation; see text).  }
  \label{f:logr_logj_mbins_meds}
\end{figure}

The cumulative specific angular momentum magnitude profiles, $\jr$
(as defined in Section~2.3), of haloes in the MS and
\hrsim{} are shown in Fig. \ref{f:logr_logj_mbins_meds}.  The vertical
error bars (directly around the data points) give an estimate of the
uncertainty in the median, by analogy to a Gaussian mean,
\begin{eqnarray}
  \epsilon_+ = \frac{X_{84}-X_{50}}{\sqrt{N}}, &  \:\: & 
  \epsilon_- = \frac{X_{50}-X_{16}}{\sqrt{N}},
  \label{e:mederr}
\end{eqnarray}
where $X_i$ is the value at the $i$th percentile of the distribution
in question, made up of $N$ objects ($X_{50}$ is the median).  These
are virtually invisible on the lines corresponding to the bulk of the 
halo populations, but quite significant in the higher mass bin medians
which contain considerably fewer haloes.  By contrast, the outer
bars and boxes (only shown on the MS median line) indicate the spread
of the data; the boxes enclose $68$ per cent of the data, and the
outer bars $95$ per cent.

Fig.~\ref{f:logr_logj_mbins_meds} illustrates the trend of $\jr$ with
mass, as well as the trend with radius at a fixed halo mass.  For
comparison, we show the radial scaling for a test particle undergoing
circular motion at radius $r$, $\jp(r) = r\vc$, where we take the
circular velocity due to the mass within $r$, $\vc$, to be constant
with radius (as for an isothermal density profile). An NFW density
profile \citep{NFW96,NFW97} deviates from an isothermal profile at
small and large $r$.  Since our haloes are well-described by NFW
profiles \citep{neto07,2008MNRAS.387..536G}, their 
angular momentum profiles should  deviate slightly from the
$\jp\propto r$ form, as indeed they do. Despite the
simplicity of these arguments, they give a good account of our
simulation results.

\begin{figure}
  \includegraphics[width=80mm]{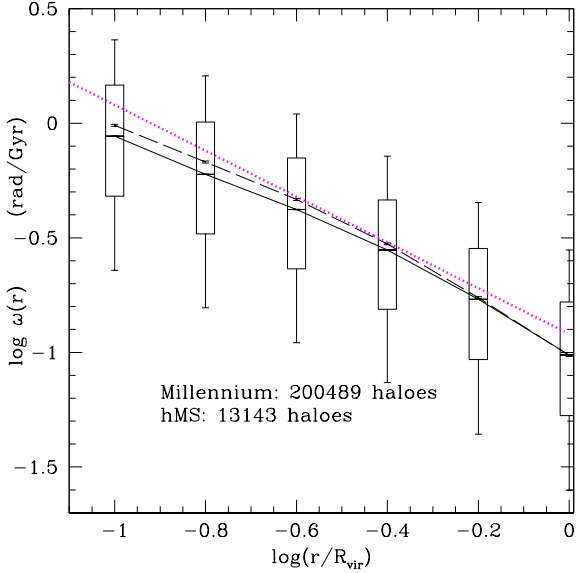} 
  \caption{Angular velocity profiles, $\omega(r)$ (see
    Eqn.~\ref{e:angvel}), for haloes in the MS (solid black) and
    \hrsim{} (dashed black) simulations.  As with
    Fig.~\ref{f:logr_logj_mbins_meds}, we plot error bars on both
    lines, and indicate the spread of the data by the outer bars and
    boxes on the MS line. The trend from dimensional arguments,
    $\omega\propto r^{-1}$, is plotted as the magenta dotted line
    (with arbitrary normalisation; see text).}
  \label{f:logr_angvel_meds}
\end{figure}

A complementary measure of the angular momentum profile is the angular
velocity profile, $\omega(r)$.  We show this in
Fig. \ref{f:logr_angvel_meds}.  We also show the simple scaling
for a test particle consistent with the $\jp\propto r$
  trend described above: since $\vc=\omega_\rmn{p}(r) r$, we have
  $\omega_\rmn{p}\propto r^{-1}$.  Using $\omega$ instead of $j$ removes
the mass dependence, but we still find a similar amount of scatter
about the median.

\cite{1987ApJ...319..575B} showed that the \emph{differential} profile
of the specific angular momentum of dark matter halos, $j(r)$, is also
roughly proportional to $r$.  \cite{2001ApJ...555..240B} considered
$j_z(r)$, the component of the (differential) specific angular
momentum at $r$ parallel to the total halo angular momentum vector;
they found that it follows $j_z(r) \propto r^\alpha$, with $\alpha =
1.1\pm 0.3$.  In the (unrealistically) simple case of perfect internal
alignment and constant circular velocity throughout the halo, we would
expect $\jr$, $j(r)$, and $j_z(r)$ to \emph{all} be proportional to
$r$.  All these studies see deviations away from this simple trend
since haloes are, of course, more complex structures; furthermore,
there is a significant amount of variation among individual haloes of
a given mass in properties such as mass profiles, spins, and internal
alignments.


\subsubsection{Spin orientation profiles}\label{s:millorient}
We now investigate the orientation of the halo angular momentum vector
and its dependence on halo mass and radius. For this, we compute the
cumulative angular momentum orientation profile,
\begin{equation}
  \cos \theta(\le r) = \hat{\vv{j}}_\rmn{inner}\dotprod\hat{\vv{j}}(\le r),
\end{equation}
where the hat denotes a unit vector (e.g. $\hat{\vv{j}} =
\vv{j}/|\vv{j}|$) and the `inner' region is now defined as $r\le
10^{-0.6}\Rvir \simeq 0.25\Rvir$.  

The direction of the cumulative angular momentum vector is subject to
significant uncertainty due to discreteness effects. To ensure that
our results are robust, we have applied a different, additional, set
of selection criteria.  The net angular momentum of a halo,
$\vv{j}(\le r)$, is constructed from the 3-D vector sum of its
individual particles' angular momenta.  But haloes, in fact, have very
little coherent rotation, so the specific angular momentum of a halo
is small compared to the typical specific angular momentum of an
individual particle.  A halo that has a particularly small $j$
compared to those of its particles will have its direction information
dominated by very few particles, which introduces a significant amount
of uncertainty.  To mitigate this problem, we restrict our sample to
haloes whose $j$ values are not too small.  We discuss the details of
this procedure in Appendix~\ref{a:cosboot} where we demonstrate that
our selection reduces the scatter and ensures that the angular
momentum directions are reliable.

\begin{figure}
  \includegraphics[width=80mm]{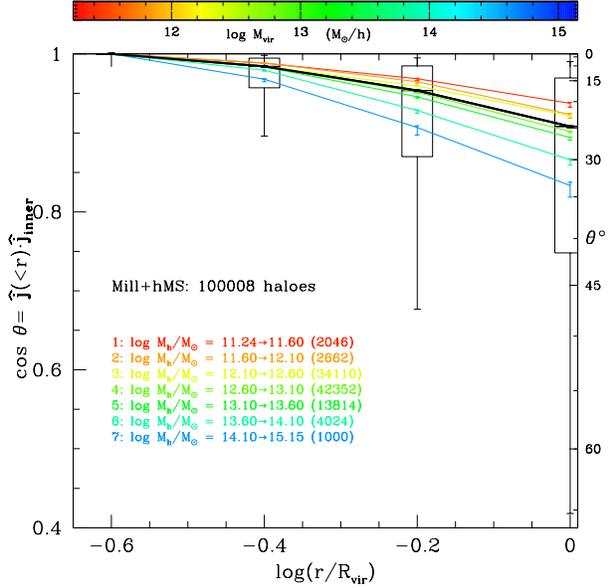} 
  \caption{Cumulative specific angular momentum orientation profiles
    for haloes from the MS and \hrsim{} simulations.  The median
    profile for the entire halo population is shown in black, with
    error bars and percentile bars shown as in
    Fig.~\ref{f:logr_logj_mbins_meds}.  There is an increasing
    likelihood of misalignment at larger radius, with a very large
    amount of scatter between haloes.  The coloured lines show the
    results for the haloes in different mass bins, showing that the
    lower mass haloes tend to remain better aligned to a larger
    radius. }
  \label{f:logr_cosjrjin}
\end{figure}

Fig. \ref{f:logr_cosjrjin} shows the orientation profiles of the
haloes from the MS and \hrsim{} simulations.  As radius increases, the
cumulative angular momentum becomes increasingly poorly aligned with
the inner region of the halo.  Although the median alignment of the
angular momentum at $\Rvir$ with that in the inner regions is always
within $30\degr$, there is a very large scatter amongst haloes.  This
scatter is much larger than that expected from the numerical
limitations discussed above, so we conclude that it reflects the
intrinsic variation amongst haloes.

\begin{figure}
  \includegraphics[width=80mm]{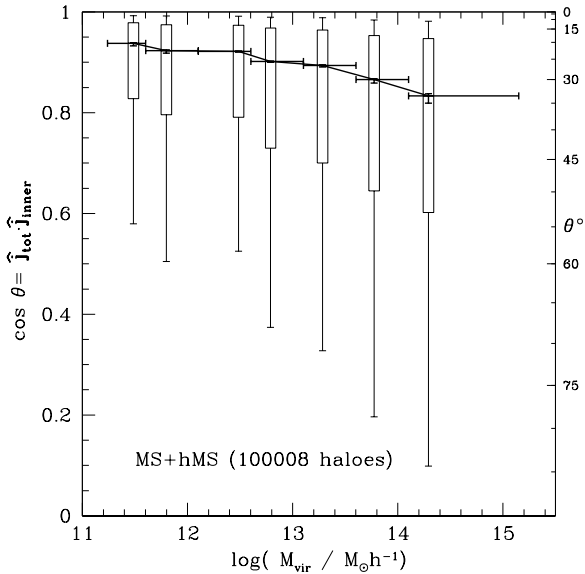} 
  \caption{Angle between the specific angular momentum within $\Rvir$
    and within the inner region ($\approx0.25\Rvir$), as a function of
    halo mass, $\Mvir$.  There is very little trend for the majority
    of haloes, but the most massive ones show a clear tendency for
    increased misalignment.  The error bars indicate the uncertainty
    on the median, and the outer bars and boxes show the spread of the
    data, as in Fig.~\ref{f:logr_logj_mbins_meds}. The horizontal bars
    indicate the mass-bin widths.}
\label{f:logm_cosjtotjin}
\end{figure}

Furthermore, when the data are split into different mass bins, we can
see that there is a clear trend, with more massive haloes tending to
be less well aligned at large radius.  Fig. \ref{f:logm_cosjtotjin}
examines the mass dependence in more detail, by showing how the angle
between $\vjtot=\vv{j}(\le\Rvir)$ and $\vjin=\vv{j}(\le 0.25\Rvir)$
varies as a function of halo mass.  While the trend is weak over a
wide range of mass at the lower end, there is a clear decrease in
alignment for the very highest mass haloes. The reason for this is
likely to be related to the
  hierarchical nature of structure formation: the most massive haloes
formed most recently, and are likely to have experienced a major
merger more recently than smaller mass haloes. Major mergers have been
found to have a strong effect on the halo's angular momentum magnitude
\citep[see e.g.][]{Frenk85,2002MNRAS.329..423M, 2002ApJ...581..799V,
  2007MNRAS.380L..58D,zavala2008}; it is highly likely that a major
  merger would decorrelate the alignment profile of the halo, albeit
  in a way that could depend strongly on the details of the merger.

\begin{figure}
  \includegraphics[width=80mm]{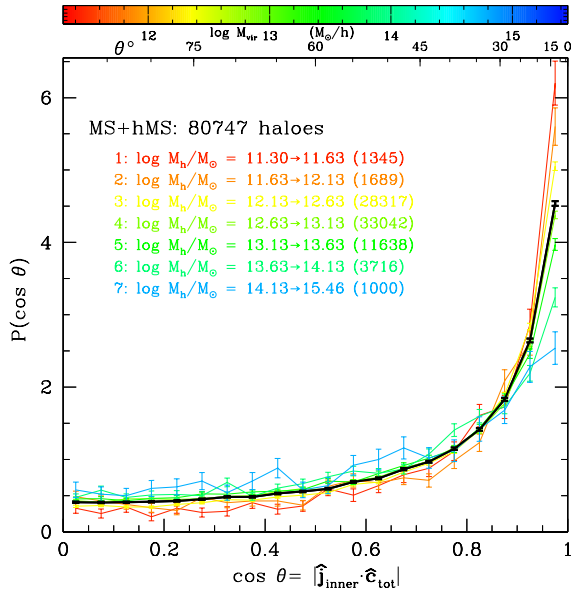} 
  \caption{Normalised histogram of the angle between the halo minor
    axis and the specific angular momentum of the halo inner region
    ($\approx0.25\Rvir$).  The overall median for the MS and \hrsim{}
    simulations is plotted in black over the coloured lines
    representing the medians in different mass bins.  The error bars
    show the Poisson uncertainty ($\sqrt{N}$) in each bin.  As well
    as excluding haloes whose $\vjin$ directions are poorly determined
    due to discreteness, we also apply an analogous selection criteria
    for the halo shape axes, eliminating the nearly-spherical objects;
    see Appendix \ref{a:cosboot} for details.}
  \label{f:cosctotjin_histo}
\end{figure}

\citetalias{Bett07} considered the alignment of the
  total halo angular momentum vector with the halo shape axes (their
  fig.~16).  Complementing that, we show the distribution of angles
  between the inner halo angular momentum and the total halo shape
  (identified by the minor axis, $\vv{c}$) in
  Fig.~\ref{f:cosctotjin_histo}.  Although the alignment tends to be
  good, there is a significant population of haloes where the
  misalignment exceeds $45\degr$.  We find a slight trend with mass
  which is also likely to arise from the difference in halo merger
  histories.  The distribution of angles between the total halo
  shape and the inner halo angular momentum is, in fact, very similar to
  that with the total halo angular momentum shown in
  \citetalias{Bett07}.

We can attempt to compare our results to those found by
\cite{BS05}, hereafter \citetalias{BS05}.  These authors compared both
the angular momentum and shape-axis vectors of haloes as a function of
radius.  At first glance, our results might seem to suggest better
internal halo alignment than found in \citetalias{BS05}. However,
the differences in halo selection and analysis make direct
comparisons complicated.  In particular, \citetalias{BS05} used the
differential angular momentum profile, $j(r)$, whereas we use the
cumulative $\jr$. Although this allows us to retain more objects, as
discussed in section \ref{s:phys}, it also smooths out quantities like
the misalignment profile, resulting in an average reduction in
misalignment compared to the differential profile. (This is discussed
in further detail in Section~\ref{s:dmodmg} where we directly compare
differential and cumulative profiles.)  While it seems likely that
this is the primary reason for the difference between the results,
there are other differences in methodology that could also play a
role.  Although the \citetalias{BS05} simulations have slightly better
mass resolution than ours, their simulations had relatively few
haloes, which could also increase the scatter compared to our results.
The halo populations in the two studies cover a very similar mass
range, and we have used similar techniques to control errors due to
numerical effects.  These include limiting the minimum number of
particles in the halo and using bootstrap resampling to test for, and
reject, haloes whose vector directions are unreliable.  In this paper,
we also explicitly remove haloes whose energies indicate that they are
out of equilibrium, although we do not find that this is as strong a
constraint as the limits on particle number and those derived from the
bootstrap analysis.  The lack of a clear trend with halo mass seen in
\citetalias{BS05} is consistent with the
results we presented in Fig.~\ref{f:logm_cosjtotjin}, since the trend
we see is quite weak and has a rather large scatter.  The similarity
in the distributions of angles between the shape axes and angular
momentum vectors at $\Rvir$ and $0.25\Rvir$ is also
consistent with
\citetalias{BS05}, given the caveats above.  Similar
  comments can be made regarding earlier work on halo internal
  alignments, such as that by \cite{1992ApJ...401..441D} and
  \cite{1992ApJ...399..405W}, both based on differential profiles. In
  so far as they can be compared directly, our results are compatible
  with these.


\section{The effect of baryons}\label{s:dmodmg}

In order to extend our results to realistic galactic haloes, it is
necessary to consider how the galaxy formation process affects the
dark matter structures.  In this section, we investigate how baryonic
process modify haloes, in comparison with  dark-matter-only
structures.

\subsection{The simulations}

We use a simulation that, at $z=0$, contains a roughly spherical,
high resolution region, with a diameter of about $12.5\lunit$,
embedded in progressively lower resolution boundary regions out to a
cubical boundary of side $35.33\lunit$.  The simulation was performed
with the modified \textsc{Gadget-2} code developed by
\cite{Takashi05}, which includes a detailed implementation of the
physical processes required for forming galaxies.

Although we briefly review these processes here, the reader is
referred to \cite{Takashi05} for full details.  The modelling of the
interstellar medium (ISM) mostly follows the method of
\cite{2003MNRAS.339..289S}: the ISM is a two phase gas, consisting of
an ambient hot phase and cool clouds, in pressure equilibrium with
each other.  The heating and cooling of gas is calculated under the
assumptions of collisional ionisation equilibrium in the presence of
a uniform ultraviolet background that evolves with time
\citep{1996ApJ...461...20H}.  Cooling depends explicitly on the gas
metallicity, using the cooling tables from \cite{1993ApJS...88..253S};
molecular cooling and metal cooling at temperatures below $10^4\,
\rmn{K}$ are not included.

Energy to heat the gas is supplied by both Type~II and Type~Ia
supernovae, which provide a feedback mechanism for the gas.  Stars form
from the cool gas in either a `quiescent' or `burst' mode.  In
quiescent star formation, stars form according to a given probability
once the cold gas density rises above some threshold.  These stars
have a \cite{Salpeter1955} initial mass function (IMF).  In 
the burst mode, stars form over a shorter timescale and with a top-heavy IMF
\citep{Baugh2005}.  This gives rise to 
stronger feedback because of the larger number of supernovae. When the
injected heating exceeds the local cooling, the hot gas can be blown
out of the galaxy.  The starbursts are triggered by the presence of
shocks, which are caused by galaxy mergers.

These prescriptions for star formation and feedback ensure that much
of the gas remains hot during the early stages of the galaxy formation
process and that, once the merger rate has subsided at later times,
the gas can cool and form stars stably, resulting in fairly realistic
galaxies.  We shall refer to the simulation ran with this code as
`DMG' (dark matter plus galaxies)\footnote{The DMG simulation was also
analysed in \cite{Libeskind07}, (their `SR' simulation) to investigate
satellite galaxy alignments.}.

In addition, we re-ran the DMG simulation without baryons -- the `DMO'
simulation -- redistributing the equivalent baryonic mass to the dark
matter particles in the initial conditions to conserve the overall
mass.  Since these two simulations have exactly the same initial mass
distribution, we can match corresponding haloes and thus directly
estimate the effects on the halo properties of interest of the
baryonic physics included in the simulations. We can then
`extrapolate' our conclusions to the MS and \hrsim{}, which lack the
complex baryonic physics but contain many orders of magnitude more
objects.

As in Section~\ref{s:mshr1}, we analyse our simulations at $z=0$.  The
assumed cosmological parameters are given in Table~\ref{t:cosparams}
(they are slightly different to those of the MS and \hrsim{}, but this
should have no impact on our results), and the simulation
parameters are given in Table~\ref{t:simparams}.

We identify haloes in DMO and DMG in the same way as in MS and
\hrsim{} (see Section~\ref{s:mshr1halodef}).  The spherical
overdensity algorithm employs Eqn.~\ref{e:virdens} to define the
`virial' radius of the haloes; for the DMO/DMG cosmology, the
overdensity parameter has the value of $\Deltac=101$.  We have to take
account of the low resolution (high mass) boundary particles that
surround the high resolution central region in the DMO and DMG
simulations.  Haloes near the edge of the high resolution region are at
risk of contamination by boundary particles.  To ensure that this does
not affect our results, we retain haloes only if there are no boundary
particles within a radius of $\Rvir+100\klunit$ of their centre.


\subsection{Galaxy identification}

We identify galaxies as collections of gas and star particles within dark
matter haloes.  Each galaxy is identified as the most massive object found by
a FOF algorithm applied to the baryonic particles (both stars and gas
together) within the virial radius of the parent halo, using a linking length
of $1.07\klunit$ (which corresponds to $b=0.02$ for the stellar particle
number density at $z=0$).  This value of $b$ was chosen to be an order of
magnitude below the value used for dark matter, since baryons collapse by a
factor of $\sim 1/(2\lambda)$, and the spin parameter $\lambda\sim0.04$
\citep[see e.g.][]{1980MNRAS.193..189F, 2001MNRAS.326..649P,
  BaughReview06}. We have checked that this identification is reasonable. The
resulting galaxies do not contain either large amounts of hot halo gas or
satellite galaxies, which would bias measurements of shape and angular
momentum.

The centre of each halo's galaxy is determined by first finding the
centre of mass of the stellar particles within the halo, and the
radius of the sphere encompassing them all.  The radius is then shrank
by $5$ per cent, and the centre of mass of the particles remaining
within the sphere is calculated.  Using the new centre, the radius is
shrank again, iterating until only $\leq 50$ particles remain.  The
galaxy centre is then taken to be the last centre of mass calculated
using at least $50$ particles.  The galaxy centres are found to
correspond very well with the halo centres: for the $99$ galaxy--halo
systems (with $Q\le 0.5$ and containing at least $1000$ particles in
both the halo and stellar component), the median galaxy--halo centre
separation, as a fraction of $\Rvir$, is $0.0058$; the mean is
$0.0064$ and the standard deviation is $0.0030$. We also define the
galaxy outer radius to be the distance from the galaxy centre to the
farthest baryonic particle included in the FOF group.

% ...
%


\subsection{Physical properties}\label{s:dmodmgprops}

The inclusion of baryons in the DMG haloes requires some definitions
to be made more carefully.  The virial masses and radii of the DMG
haloes are defined using all the mass (i.e. including baryons), as are
the kinetic and potential energies.  The smaller number of haloes in
these simulations makes it possible to compute the potential energies
using all the particles within $\Rvir$ (rather than only using 1000 as in
the MS and \hrsim{} simulations). We have to take into account the
different gravitational softening lengths and individual masses of the
baryonic particles, so the contribution to the potential energy from
each particle pair $i$--$j$ is the mean:
\begin{equation}
  u_{ij} = \frac{1}{2}
           \left(  \frac{-W_2(r_{ij}/\eta_i)}{\eta_i}
                  +\frac{-W_2(r_{ij}/\eta_j)}{\eta_j}
           \right),
\end{equation}	  
where $W_2(x)$ is the SPH smoothing kernel (see Eqns~\ref{e:pot}
and~\ref{e:softkernel}).  The total potential energy of each halo
system is therefore:
\begin{equation}
  U = -G \sum_{i=1}^{N-1} \sum_{j=i+1}^{N}  m_i m_j u_{ij}.
\end{equation}

We perform a dynamical decomposition of the stellar systems in our
simulations in order to determine the relative contributions of the
disc and bulge components, measured both according to stellar mass and
to $B$-band luminosity.  We do this for galaxies containing at least
$5000$ star particles, in the same way as was done by
\cite{Takashi05}, whose method is based on that of \cite{ANSE2003}.
The stringent particle number criterion leaves only 30 galaxies in the
sample but it ensures that we exclude galaxies whose morphologies are
biased towards bulginess due to poor numerical resolution.  We let the
angular momentum of the stellar component of each galaxy define a
`$z$'-axis, and compute the component of the angular momentum of each
star particle parallel to this direction.  Half of the bulge is
identified with the particles that have $j_z<0$; the total bulge mass
is defined as twice the mass of those particles.  The disc mass is
then given by the difference between the total stellar mass and the
bulge mass.  Using the same method, we also compute the disc-to-total
ratio ($D/T$) in terms of the star particles' $B$-band luminosity (see
\cite{Takashi05} for details).

\begin{figure}
  \centering\includegraphics[width=80mm]{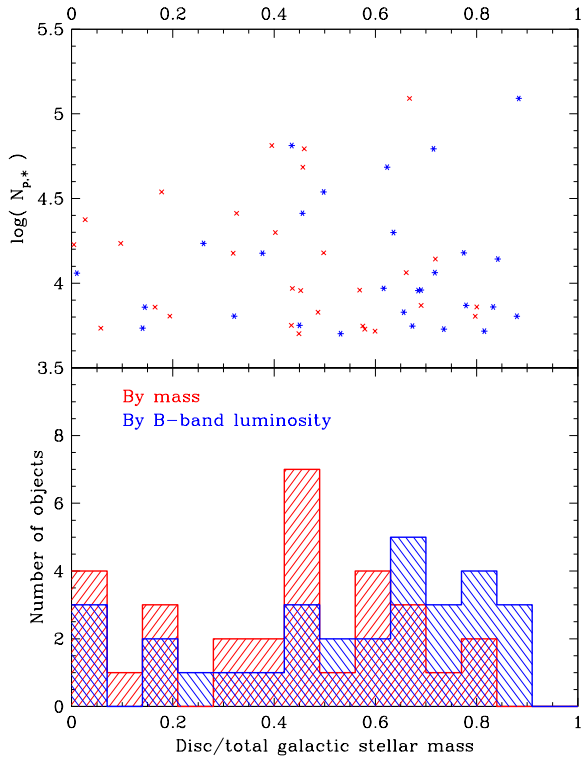} 
  \caption{Disc-to-total ratios ($D/T$) for galaxies containing at
    least $5000$ star particles (absolute $B$-band magnitude brighter
    than $-17.9$), in haloes containing at least $1000$ particles, as
    a function of the number of star particles in each galaxy.  The
    upper panel shows the number of particles as a function of $D/T$
    and the lower panel shows the distribution of this quantity.  We
    plot results for the $D/T$ ratio according to stellar mass (red
    histogram and crosses) and according to $B$-band luminosity (blue
    histogram and asterisks).  The constraint on the number of star
    particles reduces the sample to $30$ galaxies, but their $D/T$
    values do not depend significantly on the number of
    particles. Imposing the virial ratio limit $Q\le 0.5$ has no
    effect on the selected sample.}
\label{f:d2tdistro}
\end{figure}

The distribution of $D/T$ ratios for galaxies in our simulations is
shown in Fig.~\ref{f:d2tdistro}. The faintest galaxy in the sample has
a total $B$-band luminosity corresponding to an absolute magnitude of
$-17.9$. This distribution does not depend significantly on the number
of star particles in the galaxy. Two thirds of our sample have
$B$-band $D/T$ ratio greater than 0.5. Thus, our simulation produces a
distribution of morphological types that is broadly consistent with
observations \citep{2007MNRAS.379..841B,2009MNRAS.396.1972P}. This is
a significant success of our simulations.

\subsection{Results}
The basic halo selection for the DMO and DMG haloes is the same as for
MS and \hrsim{} described in Section~\ref{s:milhalosel}: we require
haloes to have $Q\le 0.5$, and $\Np(\le r)\ge 300$ within the
innermost radius considered. Using $0.1\Rvir$ as the innermost radius
yields $83$ haloes from DMO and $88$ haloes from DMG.  Again, we use
slightly different selection criteria for our investigation of angular
momentum orientations (see Appendix \ref{a:cosboot}).

\begin{figure}
  \centering\includegraphics[width=80mm]{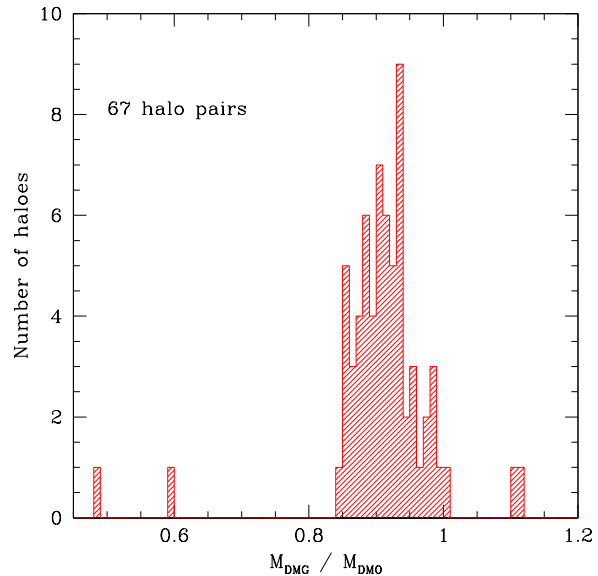}  
  \caption{Distribution of the mass ratios of matched haloes
      in the DMO and DMG simulations.  The haloes in each simulation
      are selected to have $Q\le 0.5$, and at least $300$ dark matter
      particles within $0.1\Rvir$.  Matching the haloes in the two
      simulations involves finding each halo's closest counterpart
      within $100\klunit$.  The vast majority of matched halo pairs
      have very similar masses to each other.}
  \label{f:matchmassratio}
\end{figure}

In order to compare the same haloes in the simulations with and
without baryons, we match each selected dark matter halo in DMG with
its counterpart in the set of selected DMO haloes, by finding the
closest DMO halo centre within $100\klunit$ of the DMG halo centre.
Out of the $83$ and $88$ haloes selected as described above, clear
matches are found for $67$ halo pairs.  Fig.~\ref{f:matchmassratio}
shows the distribution of their mass ratios.  The few unmatched haloes
do not have a counterpart within $100\klunit$ that passes the
selection criteria.  This failure to match all of the haloes is a
feature of the simulations, not of the matching algorithm -- it does
not depend on the selection criteria or the limiting separation.
Indeed, the fact that the same selection criteria yield differing halo
numbers for the DMO and DMG simulations hints at this.  This
difference is likely to be due to the baryons having a significant
influence on the evolution of some haloes, causing some to merge in
one simulation but not in the other, or ocassionally causing some
haloes to have very different numbers of dark matter particles.

\subsubsection{Angular momentum profiles}

\begin{figure}
  \includegraphics[width=80mm]{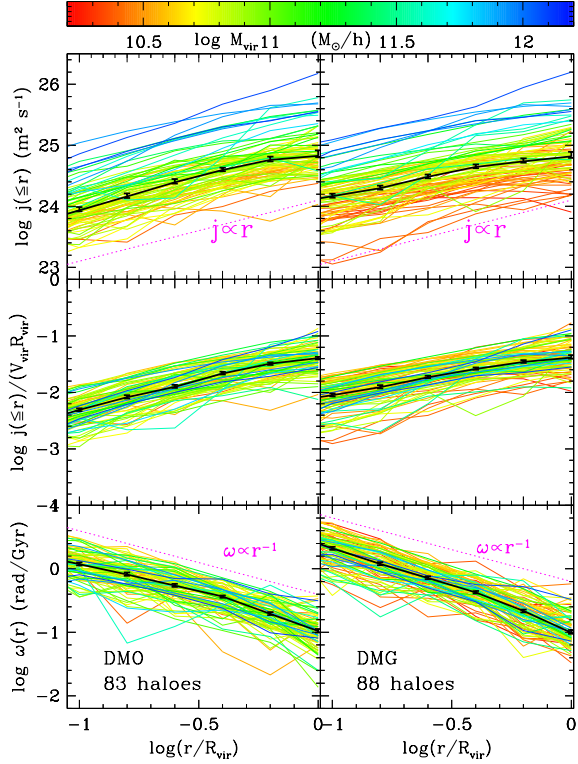} 
  \caption{Cumulative radial profiles of: specific angular momentum
    (top), scaled angular momentum $\jr/\left(V_\rmn{vir}\Rvir\right)
    = \jr/\sqrt{G\Mvir\Rvir}$ (middle) and angular velocity (bottom),
    for the dark matter in haloes from the DMO and DMG simulations
    (left and right panels respectively).  The profile for each halo
    is colour coded by its total mass ($\Mvir$).  The heavy black
    lines are the median profiles, with error bars calculated
    according to Eqn.~\ref{e:mederr}.  As a guide, the dotted magenta
    lines show the radial scaling derived from a simple argument
    assuming constant circular velocity, with arbitrary normalisation
    (see text).  Scaling the angular momentum of each halo by its
    circular velocity and radius removes the mass dependence of
    $\jr$. }
  \label{f:logr_logjmulti_dmodmg}
\end{figure}

We show the cumulative specific angular momentum profiles of haloes in
the DMO and DMG simulations in Fig.~\ref{f:logr_logjmulti_dmodmg}.
These show the same basic trends that we saw in the MS and \hrsim{}
simulations (Figs.~\ref{f:logr_logj_mbins_meds} and
\ref{f:logr_angvel_meds}; we also plot the same $j\propto r$ line for
reference).  The mass dependence of $\jr$, present in the top panels,
is removed by scaling by $\Vvir\Rvir \equiv \vc(\Rvir)\Rvir$, where
the circular velocity $\vc(r) = \sqrt{GM(\le r) r^{-1}}$. We plot $\jr
/ \left(\Vvir\Rvir\right)$ in the middle panels of
Fig.~\ref{f:logr_logjmulti_dmodmg}. The mass dependence is also absent
when plotting the angular velocity, $\omega(r)$.  There is a similar
degree of halo-to-halo scatter as in the MS and \hrsim{} simulations.

\begin{figure}
  \includegraphics[width=80mm]{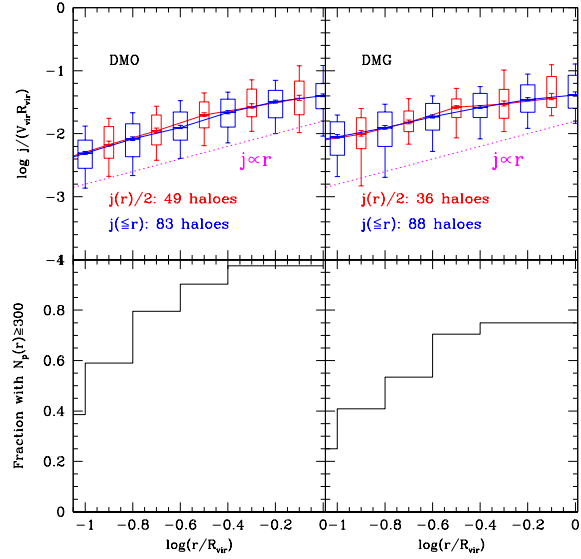} 
  \caption{Comparison of the cumulative and differential specific angular
    momentum profiles.  In the top panels, we plot the median of the
    cumulative $\log_{10} \jr/(\Vvir\Rvir)$ (blue; same as the middle panels
    of Fig.~\ref{f:logr_logjmulti_dmodmg}), and the differential $\log_{10}
    j(r)/(2\Vvir\Rvir)$ (red; see text).  The differential data are plotted at
    the centres of their radial bins, with the cumulative data plotted at the
    upper ends.  For the differential quantities, the bin that ends at
    $0.1\Rvir$ has the same width in $\log_{10} r$ as the other bins, rather
    than encompassing all mass down to the centre.  To illustrate the scatter
    in the data, we plot boxes and outer bars enclosing the $68$ and $95$ per
    cent of the distributions in each bin respectively.  The lower panels show
    the fraction of haloes in each radial bin that have at least $300$ dark
    matter particles, relative to the number selected for the cumulative
    calculation. }
\label{f:logr_logjmultiNp_dmodmg_noncuml}
\end{figure}

For completeness, and to aid comparisons, we have also computed the
\emph{differential} specific angular momentum profiles for the DMO and DMG
haloes.  The cumulative and differential specific angular momenta are
related through: 
\begin{equation} 
\jr = \frac{\int_0^r j(r')\rho(r') {r'}^2 \dd r'}
{\int_0^r \rho(r') {r'}^2 \dd r'} \label{e:jdiffdfn}.
\end{equation} 
If $j(r)\propto r$, and the density profile $\rho(r)\propto r^{-2}$
(at least, for the radius and mass ranges of interest), then the
cumulative specific angular momentum is simply $\jr = j(r)/2$.  In
Fig.~\ref{f:logr_logjmultiNp_dmodmg_noncuml}, we plot the medians of
both $\jr$ and $j(r)/2$, scaled by $\Vvir\Rvir$ (as in the middle
panels of Fig.~\ref{f:logr_logjmulti_dmodmg}) to remove the dependency
on halo mass.  The cumulative and differential angular momentum
profiles are very similar, but there are far fewer halos included in
the differential calculation.

Our selection criteria for accurate angular momentum measurements
requires that, in the differential case, \emph{each} bin should
satisfy the conditions described in Section~\ref{s:milhalosel}.  In
particular, each bin has to contain at least $300$ dark matter
particles.  In the lower panels of
Fig.~\ref{f:logr_logjmultiNp_dmodmg_noncuml}, we show the fraction of
the $83$ and $88$ halos selected for the cumulative calculation that
are also selected for the differential calculation, i.e.  that have at
least $300$ dark matter particles in each radial bin.  Because of the
relative lack of material in the inner bins, there is a
significant reduction in the number of selected haloes in the
differential case (of course, preferentialy excluding those with lower
masses).

\begin{figure}
  \includegraphics[width=80mm]{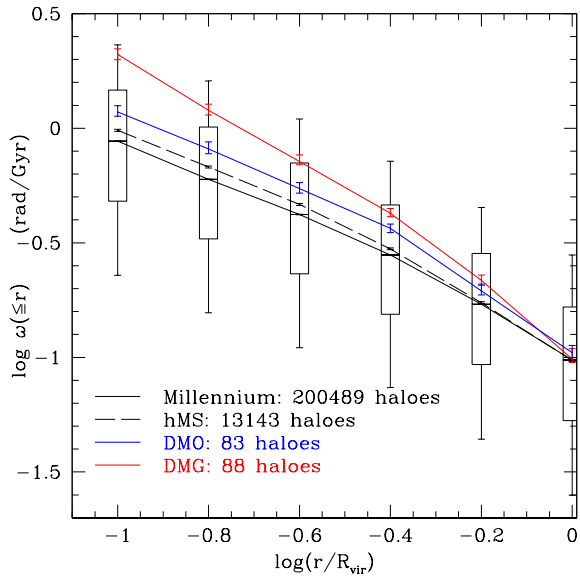} 
  \caption{The median dark matter angular velocity profiles for haloes
    in our four simulations.  The profiles for haloes containing
    baryons (DMG) have a significantly greater angular velocity in
    their inner regions, in the median.  As in previous figures, the
    error bars give the uncertainty in the median
    (Eqn.~\ref{e:mederr}) and the outer bars and boxes plotted on the
    MS curve give the spread of the data (the boxes enclose $68$ per
    cent, and the outer bars $95$ per cent of the data). }
  \label{f:logr_angvel_medsall}
\end{figure}

\begin{figure}
  \includegraphics[width=80mm]{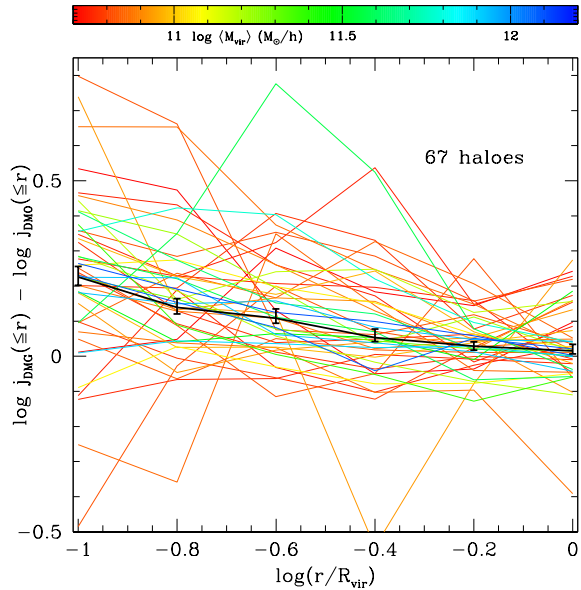} 
  \caption{Logarithm of the ratio of the cumulative specific dark
    matter angular momentum profiles of haloes from the DMG simulation
    to that of the matched haloes in the DMO simulation.  Each line
    represents the ratio from a matched DMG--DMO halo pair, colour
    coded according to the mean mass of the two haloes.  The median
    ratios are marked by the heavy black line, with error bars given
    according to Eqn.~\ref{e:mederr}. }
  \label{f:logr_dmodmgmatchrat}
\end{figure}

Fig. \ref{f:logr_angvel_medsall} shows the median value of
$\omega(r)$, the angular velocity at radius $r$ due to the mass
internal to that radius, in each of our four simulations. It is
immediately apparent that the dark matter in the inner regions of
haloes containing galaxies spins significantly faster than in their
dark matter-only counterparts.  This can be seen even more clearly in
Fig.~\ref{f:logr_dmodmgmatchrat}, where we plot the ratio of $\jr$ for
each DMG halo to that of its DMO counterpart. Note that we have fewer
halos after matching, as discussed above. (The equivalent plot for
ratios of $\omega(r)$ is very similar.)  There is a large halo-to-halo
scatter but, at the virial radius, the dark matter remains unaffected
by the baryons (in the median).  However, within $\Rvir$, the trend in
the median is for the angular momentum of the inner 10 per cent of the
dark matter in a halo to be $\sim 50$ per cent greater if that halo
has a galaxy in it than if it does not.

There are two possible reasons for the difference between the $\jin$
of the DMG and DMO haloes. The addition of baryons can cause the dark
matter to contract adiabatically, so that a given mass is contained
within a smaller radius, and therefore rotates faster.  In this case,
the angular momentum of the dark matter would be conserved
independently of that of the baryons.  Alternatively, the dark matter
could also gain angular momentum through the transfer from the
infalling gas by tidal torques and dynamical friction. We now look
into these possibilities in more detail.

We expect a DMG halo to be more centrally concentrated than its DMO
counterpart because its baryons will have dissipated their energy,
fallen to the centre and deepened the potential well (see e.g.
\citealt{1984MNRAS.211..753B}, \citealt{1986ApJ...301...27B},
\citealt{gnedin2004}). Dark matter will fall into the deeper potential
well and, if it conserves its own angular momentum, it would spin up,
giving rise to an increase in $\jr$ in the central regions. Thus, if
the angular momentum of the dark matter of a DMG halo within a given
inner radius is the same as that of the dark matter of its DMO
counterpart within some \emph{larger} radius, which contains the same
dark matter mass, then there has been no net transfer of angular
momentum between the dark matter and the baryons: the dark matter has
simply contracted (perhaps adiabatically).

\begin{figure}
  \includegraphics[width=80mm]{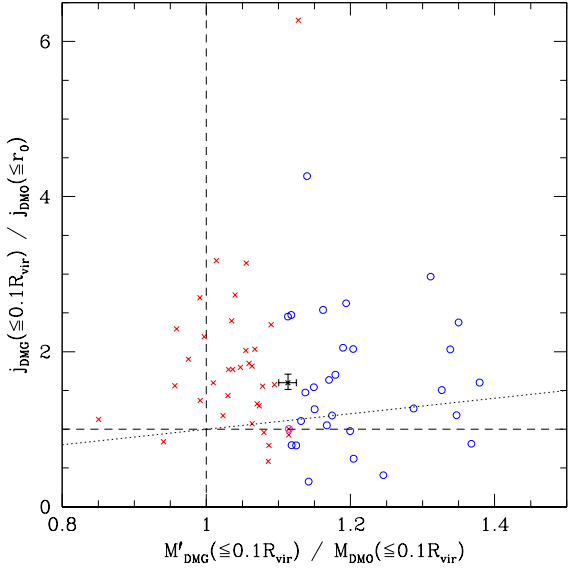} 
  \caption{Ratio of the specific angular momentum of the dark matter
    mass within $0.1\Rvir$ of each DMG halo, to that of the
    corresponding DMO halo at a radius $r_0$ that contains the same
    mass.  A value of unity (dashed horizontal line) indicates that
    the dark matter has conserved its own angular momentum.  This
    ratio is plotted against the ratio of dark matter masses at
    $0.1\Rvir$ for each halo pair. (The DMG dark matter mass is scaled
    by the baryon fraction, see text).  Values greater than unity
    (dashed vertical line) in this axis indicate that the halo has
    contracted. To highlight the relative scale, the 1:1 relation is
    marked with a dotted line.  Haloes are selected according to our
    usual criteria of $\Np({\le 0.1\Rvir})\ge 300$ and $Q\le 0.5$.
    Blue circles mark the $29$ halo pairs with $r_0\ge 0.1\Rvir$; red
    crosses mark the $34$ haloes where $r_0<0.1\Rvir$.  In the latter
    case, the measured $j({\le r_0})$ could come from fewer than $300$
    particles, but in practice this only occurs for one halo (marked
    with a magenta circle around its red cross) since most have
    $r_0\approx 0.1\Rvir$.  The median of all the halo pairs excluding
    this one is marked with a black cross, with error bars given by
    Eqn.~\ref{e:mederr}. }
  \label{f:massjrtests}
\end{figure}

Let $M^\prime_\rmn{DMG}$ be the mass of dark matter contained within
$0.1\Rvir$ of the DMG halo, appropriately scaled by the baryon
fraction\footnote{When comparing the dark matter mass between the DMO
  and DMG haloes, we scale each DMG halo dark matter mass by $\bfrac$,
  the fraction of mass in baryons within $\Rvir$; i.e. we set
  $M^\prime(\le r) = M(\le r)/\left(1-\bfrac\right)$.}, and let $r_0$
be the radius in the corresponding DMO halo that contains the same
mass, i.e:
\begin{equation}
  M^\prime_\rmn{DMG}(\le 0.1\Rvir) = M_\rmn{DMO}(\le r_0)
\end{equation}
(If the dark matter has contracted substantially, we would
  expect $M_\rmn{DMO}({\le r_0}) > M_\rmn{DMO}({\le 0.1\Rvir})$.) We
then compute the ratio between the dark matter angular momentum,
 $j_\rmn{DMG}({\le 0.1\Rvir})$, and $j_\rmn{DMO}({\le r_0})$; if the
  dark matter angular momentum has been conserved, this should be
unity. In Fig.~\ref{f:massjrtests}, we compare
this to the ratio of dark matter masses within $0.1\Rvir$. (We find
that, in practice, $R_\rmn{vir,DMG}\simeq R_\rmn{vir,DMO}$, so we do
not distinguish them in this plot.)
Fig. \ref{f:massjrtests} shows that the increase in angular momentum
of the inner dark matter in DMG haloes at fixed mass (median of $60$
per cent) is much greater than the corresponding change in mass at
that radius (median value of $10$ per cent). The fact that $j$ grows
even at fixed dark matter mass suggests that the angular momentum of
the dark matter is not conserved, but instead gains, at least in part,
from the baryons.  This is consistent with the results of
\cite{Kaufmann2007}, who performed a detailed investigation of the
different ways in which angular momentum can be transported away from
the cooling gas in a halo, given that current simulations have not yet
reached the resolution whereby the final angular momentum of the gas
has converged.

Our work builds upon the results of \cite*{zavala2008}, which was also
based on simulations by \cite{Takashi05}. They found that in a
simulation with weak star formation and feedback, a large amount of
angular momentum was transferred from the baryons to the dark matter,
resulting in a bulge-dominated galaxy. On the other hand, in a
simulation from {\em the same initial conditions} but stronger
feedback, the baryons approximately conserved their angular momentum
resulting in a disc-dominated galaxy.  The simulations that we analyse
here use the same strong feedback model. Our results for a sample of
many objects indicate that, in fact, some angular momentum transfer
still takes place on average even in the strong feedback case. The
size of this transfer, however, is sufficiently small and has
sufficiently large scatter that the majority of the galaxies still end
up being disc-dominated.


\subsubsection{Spin orientation profiles}\label{s:dmodmgorient}
\begin{figure}
  \includegraphics[width=80mm]{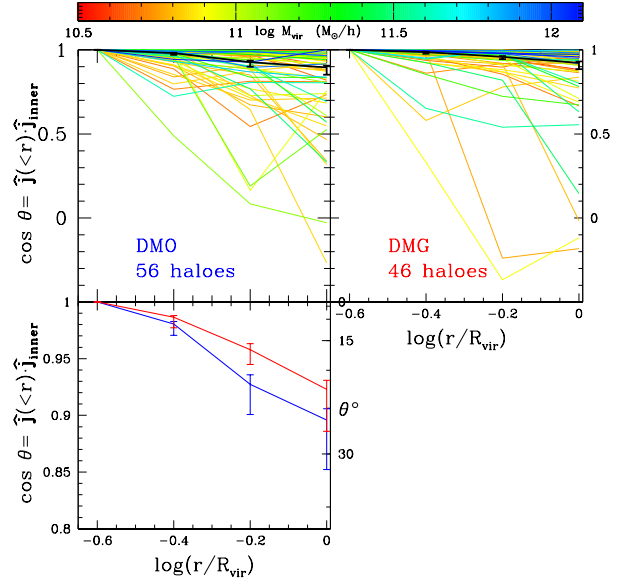} 
  \caption{Cumulative dark matter angular momentum orientation
    profiles of haloes in the DMO and DMG simulations (top left and
    right respectively).  Each halo is colour coded according to its
    mass, with the median profile shown in heavy black, and error bars
    plotted according to Eqn.~\ref{e:mederr}.  The bottom-left panel
    re-plots just the two median lines and their error bars, with DMO
    and DMG shown in blue and red respectively.  }
  \label{f:logr_cosjrjin_dmodmg}
\end{figure}

We examine the dark matter cumulative spin orientation profiles of
haloes in the DMO and DMG simulations in the same way as we did
earlier for the MS and \hrsim{} data; note that the `inner' region is
again set at $\approx0.25\Rvir$.  From the individual halo profiles
shown in Fig.~\ref{f:logr_cosjrjin_dmodmg}, we can see that, as with
the MS and \hrsim{} haloes, there is a trend in the median such that
the dark matter angular momentum vector at $\Rvir$ is $15$--$30\degr$
away from that of the inner dark matter, and also that there is a very
large scatter about that trend.  There is no discernible trend with
halo mass; given the small mass range of these haloes, however, this
is consistent with the results shown earlier
(Fig.~\ref{f:logm_cosjtotjin}). When comparing the median trends of
DMO to those of DMG (lower panel of
Fig.~\ref{f:logr_cosjrjin_dmodmg}), there is a suggestion that the
haloes that have experienced baryonic physics have a total spin that
is slightly more closely aligned than the dark-matter-only haloes.
However, the two curves are within each others' error bars, so this
result on its own is inconclusive.

We have also computed the \emph{differential} orientation profiles of
the dark matter angular momentum.  As for the differential angular
momentum magnitude profiles
(Fig.~\ref{f:logr_logjmultiNp_dmodmg_noncuml}), the results are
qualitatively similar to the cumulative case, but there is now much
more scatter.  This is due both to the fact that the points in each
bin are now independent, and to the reduction in the number of haloes
selected. The restriction that each bin must contain at least $1000$
dark matter particles leaves us with just $21$ and $18$ haloes from
the DMO and DMG simulations respectively.  The median misalignment
between the inner dark matter and that at $\Rvir$ is increased to
$30$--$35\degr$.

\begin{figure}
  \includegraphics[width=80mm]{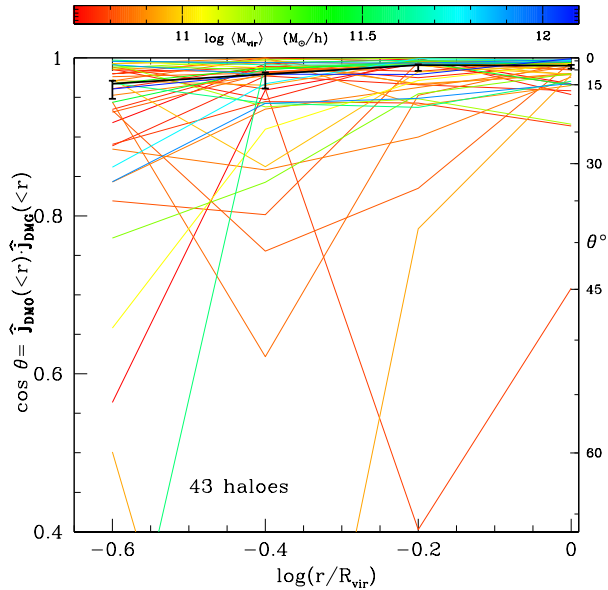} 
  \caption{Angle between the cumulative specific angular momentum
    vectors of dark matter in the DMG simulation and their
    counterparts in the DMO simulation. (A value of $\cos \theta = 1$
    means that the dark matter mass within that radius in the DMG halo
    is aligned with that in the DMO simulation.)  Each halo pair is
    colour coded according to the mean of the two halo masses, and the
    median trend with error bars is marked with the heavy black line.
  } \label{f:logr_cosjdmodmg}
\end{figure}

In order to examine possible differences between their alignment
distributions in more detail, we compare the cumulative orientation
profiles of DMG haloes directly with their counterparts in the DMO
simulation.  These results are shown in Fig.~\ref{f:logr_cosjdmodmg}.
There is a definite tendency for the baryonic processes inside the DMG
haloes to change the orientation of the inner dark matter angular
momentum, while the outer regions of the haloes remain well aligned
with their DMO counterparts.  We conclude that the baryons tend to
cause the inner regions of haloes to become better aligned with their
total halo angular momentum vector. This conclusion is reinforced by
the tentative results from Fig.~\ref{f:logr_cosjrjin_dmodmg}.

This result is in agreement with that of \cite{Bailin05} who
calculated the angle between the minor axis of haloes simulated with
and without galaxy formation physics.  They found that the inclusion
of baryonic processes reorients the inner halo shape axes while leaving
the outer halo unchanged.


\subsubsection{Dark matter--galaxy alignment distributions}
For most practical purposes, the orientation of a halo relative to its
galaxy is more relevant than the intrahalo orientation profile itself.
The alignment of a galaxy with its halo is, in principle, an
observable quantity, since gravitational lensing of background objects
can be used to measure the size and shape of the mass distribution
surrounding a galaxy (although in practice this is extremely
difficult).  The galaxy--halo orientation is also an important input
to semianalytic models of galaxy formation which, with this
information in hand, could be used, for example, to construct mock
galaxy catalogues relevant to lensing studies.

As shown in previous work (on these same simulations,
\citealt{Libeskind07}), the galaxy angular momentum is a very accurate
proxy for the orientation of the galaxy itself (i.e. the minor axis of
its mass distribution)\footnote{We find the median cosine of the angle
  between the minor axis and angular momentum axis of the stellar
  components of a galaxy is $0.9949^{+0.00047}_{-0.00443}$. (This uses
  the $81$ galaxies with $\ge 1000$ stellar particles, and stellar
  shape axis ratios of $s=c/a\leq 0.81$, in haloes with $Q\le 0.5$ and
  at least $1000$ dark matter particles.)}.  Therefore, in this
section, we shall use the angular momentum of the stellar component of
the galaxies ($\vv{j}_\rmn{gal}$), in addition to the minor axis of
the stellar mass distribution ($\vv{c}_\rmn{gal}$), to define their
orientation.

\begin{figure}
  \includegraphics[width=80mm]{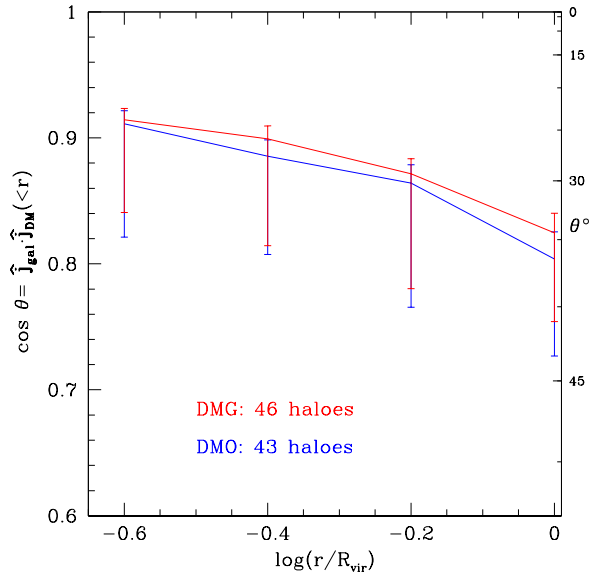} 
  \caption{Orientation profiles of the galaxies in the DMG haloes with
    respect to the cumulative dark matter angular momentum vector of
    the DMG parent haloes (red), and the corresponding DMO haloes
    (blue).  Only the median profiles are plotted.  Note that the
    error bars are the uncertainties on the median
    (Eqn.~\ref{e:mederr}), \emph{not} the spread of the data, which is
    much larger. } \label{f:logr_cosjgaljdmodmg}
\end{figure}

Fig.~\ref{f:logr_cosjgaljdmodmg} shows the median radial profile of
the orientation of the galaxies in the DMG simulation with respect to
the angular momentum vector of the dark matter of either their parent
halo in DMG, or of the corresponding halo in DMO. The scatter in the
angles is very large and so we do not show the usual percentile boxes
and bars; instead the error bars give the uncertainty in the median.
Only a rather weak trend with radius is apparent: the median values
increase from $25\degr$ at $\approx0.25\Rvir$ to $35\degr$ at the
virial radius, but given the large scatter, this is of very low
significance.

\begin{figure}
  \includegraphics[width=80mm]{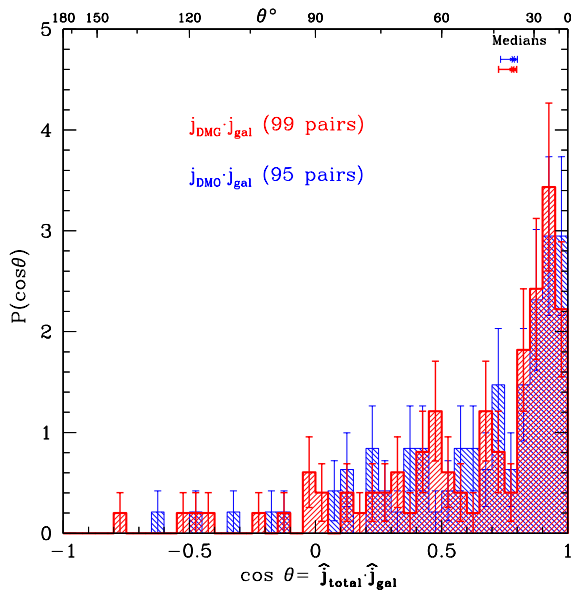} 
  \caption{Distribution of the angle between the specific angular
    momentum vector of the galaxies in DMG haloes and the specific
    angular momentum of the total dark matter of the DMG halo itself
    (red) or the corresponding DMO halo (blue).  The medians of the
    distribution are marked (at arbitrary heights) with error bars
    given by Eqn.~\ref{e:mederr}.}
  \label{f:cosjdkmjgal_histo}
\end{figure}

\begin{figure}
  \includegraphics[width=80mm]{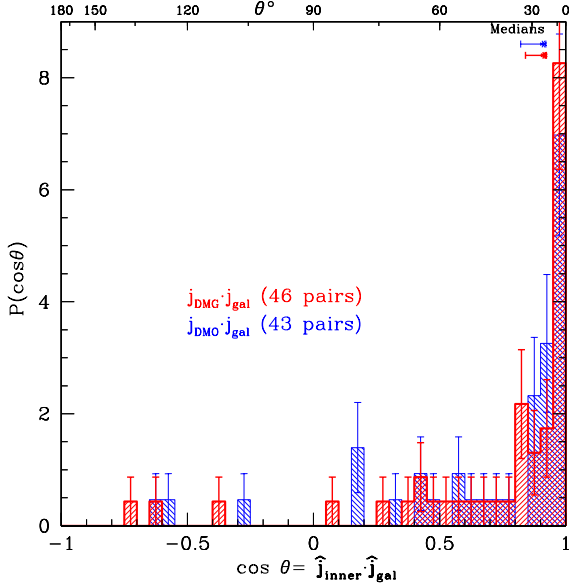} 
  \caption{As Fig.~\ref{f:cosjdkmjgal_histo}, but this time for the
    angle between the \emph{inner} halo dark matter angular momentum
    vector and the central galaxy.  The medians and their
    uncertainties are marked. }
  \label{f:cosjinnjgal_histo}
\end{figure}

\begin{figure}
  \includegraphics[width=80mm]{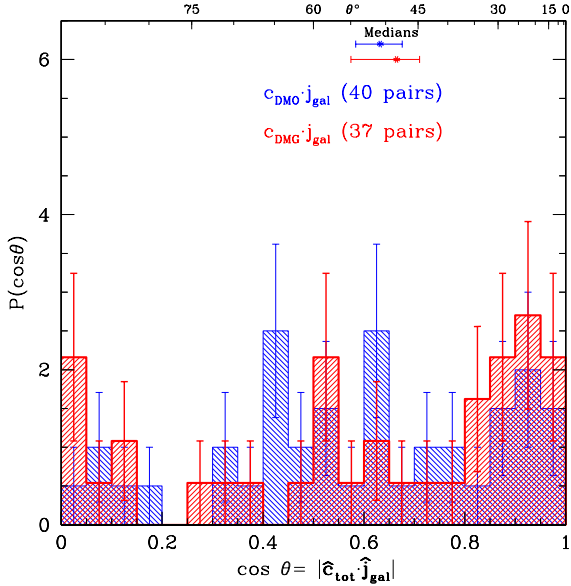} 
  \caption{As Fig.~\ref{f:cosjdkmjgal_histo}, but for the angle
    between the halo minor axis $\vv{c}_\rmn{tot}$ and the spin axis
    of the central galaxy.  Since there is no distinction between
    parallel and antiparallel for the halo shape axis, we plot the
    absolute value of the dot product. }
  \label{f:coscdkmjgal_histo}
\end{figure}

\begin{figure}
  \includegraphics[width=80mm]{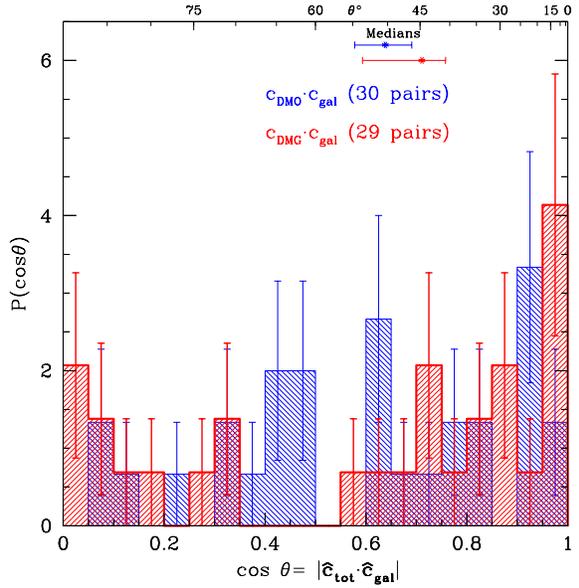} 
  \caption{As Fig.~\ref{f:coscdkmjgal_histo}, but for the angle
    between the minor axis of the halo $\vv{c}_\rmn{tot}$, and that of
    the central galaxy $\vv{c}_\rmn{gal}$.  There is no selection by
    $\jsc$, but galaxies are selected by their axis ratio, taking the
    same limit as the halo shape.}
  \label{f:coscdkmcgal_histo}
\end{figure}

\begin{figure}
  \includegraphics[width=80mm]{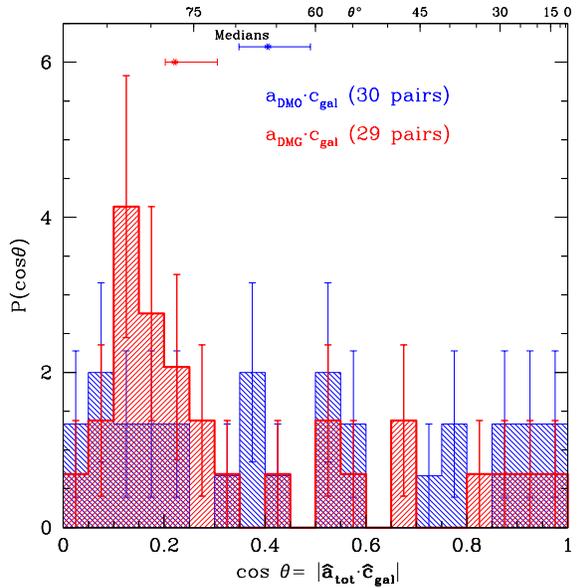} 
  \caption{As Fig.~\ref{f:coscdkmcgal_histo}, but for the angle
    between the halo major axis $\vv{a}_\rmn{tot}$ and the minor axis of the
    central galaxy.}
  \label{f:cosadkmcgal_histo}
\end{figure}

We examine the angular momentum alignment distributions themselves in
Figs.~\ref{f:cosjdkmjgal_histo} and~\ref{f:cosjinnjgal_histo}. These
show histograms of the cosine of the angle between $\vv{j}_\rmn{gal}$
and the dark matter $\vjtot$ and $\vjin$ respectively.  There is a
clear concentration towards small angles, particularly for the inner
halo, but both distributions have a long tail towards large
angles. Furthermore, the distributions are very similar for the DMG
and DMO haloes. This relatively weak galaxy--halo alignment serves to
wash out the subtle changes in the relative orientations of the dark
matter from the DMO and DMG simulations seen in the previous section.

Fig.~\ref{f:coscdkmjgal_histo} shows the histogram of the angle
between the full halo minor axis, $\vv{c}_\rmn{tot}$, and the galactic
spin axis.  In this case, the scatter between the orientation of the
halo angular momentum and its mass distribution
(Fig.~\ref{f:cosctotjin_histo} and \citetalias{Bett07}) tends to wash
out most of the (already weak) alignment between galaxy and halo.
Finally, we consider the galaxy--halo misalignment purely in terms of
the shape of the system: we plot the distribution of the angle between
the halo minor axis and the galactic stellar minor axis in
Fig.~\ref{f:coscdkmcgal_histo}.  (Although we no longer need to select
by the angular momentum magnitude, we do have to exclude galaxies that
are overly spherical, just as we do for the haloes, which reduces our
sample size further.)  The alignment distribution is still very broad,
but there is a slightly more pronounced peak towards alignment.  A
large number of galaxies are oriented perpendicular to the halo and
half of them are misaligned by $45\degr$ or more.  In
Fig.~\ref{f:cosadkmcgal_histo}, we show the distribution of the angle
between the galaxy and the \emph{major} axis of the halo
($\vv{a}_\rmn{tot}$).  About $10$ per cent of the galaxies have their
minor axes aligned within $30\degr$ of their halo's major axis.  In
the next subsection, we will explore the consequences of these results
for gravitational lensing studies.

We have also examined the alignment distributions when splitting the
galaxy populations according to their disc-to-total ratios (see
Section~\ref{s:dmodmgprops}). The resulting galaxy samples are small,
however, and we find no significant dependence of galaxy--halo
alignment on galaxy morphology.

In their analogous work on halo shape axes, \cite{Bailin05} found that
the presence of baryons forced the inner halo to align itself with the
galactic disc.  We do not find such a strong correlation. This may be
due to a combination of factors including the different physical
models adopted in the two studies and, importantly, the larger sample
size in our simulations. Our results are consistent with those
obtained for a sample of four haloes by \cite{2006PhRvD..74l3522G},
both as regards the variability from object to object and the overall
trend.

\cite{Croft2009} measured the alignments between shape and angular
momentum for the different components of their galaxies (at $z=1$),
defined as the dark matter and the gas and stars in the self-bound
subhaloes in their simulation.  Their results show slightly worse
alignment than we find (median angle of $43.5\degr$ between dark
matter and stars, compared with our values of $23.9\degr$ and
$34.4\degr$ for the median angle between the galaxy and the inner and
total halo respectively). The difference in the type of objects
considered in each study is significant; we only consider the central
galaxy in each halo, as our simulations do not have enough resolution
to study subhaloes properly.  We also take great care to remove
objects whose combination of angular momentum size and particle count
would lead to large errors in the determination of
orientation. Including such objects would have introduced greater
scatter in the results. \cite{Croft2009} do perform resolution tests,
however, which are very informative. In their lower resolution run,
the median alignment is significantly poorer.  The resolution in our
simulations is slightly better than theirs, so our results could be
consistent.

In recent years, studies based on the SDSS and 2dFGRS have shown that
satellite galaxies in haloes are preferentially aligned along the
major axis of their central galaxy \citep{2005ApJ...628L.101B,
  2006MNRAS.369.1293Y, 2007MNRAS.376L..43A, 2008MNRAS.385.1511W,
  2004MNRAS.348.1236S, 2009MNRAS.395.1184S}.  By comparing with
simulations, \cite{2006ApJ...650..550A} and \cite{2007MNRAS.378.1531K}
showed that if the central galaxies were oriented such that their
angular momentum axis were aligned with their halo's minor axis or
angular momentum axis, then the resulting satellite distribution would
be preferentially aligned with the central galaxy's major axis.  In
fact, the latter paper showed that if the galaxy and halo minor axes
were perfectly aligned, then the resulting satellite alignment signal
would be much stronger than observed. A misalignment by $\sim 40\degr$
on average, such as is produced by orienting the galaxy minor axis
with the halo angular momentum, is required to bring the signal down
to match the observations. \cite{2008MNRAS.385.1511W} have recently
shown that this result applies to haloes with masses almost down to
those in our DMO/DMG simulations.  Fig.~\ref{f:coscdkmcgal_histo}
shows that our galaxy--halo (mis)alignments are fully consistent with
the picture presented in these studies, with a median angle beween
galaxy and halo minor axis of $\sim 45\degr$.  \cite{Libeskind07} used
this DMG simulation to study the alignment of satellite galaxies, and
found results consistent with the observations, although there were
only three sufficiently well-resolved systems of satellites to study.


\subsubsection{Alignment of the projected mass distribution}

The misalignment between haloes and their galaxies can have very
important consequences for observational attempts to measure halo
properties, particularly from gravitational lensing data.  In
practice, the lensing signal from an individual galaxy halo is
too weak to be useful; the lensing distortion has to be averaged over
many galaxies by stacking data from appropriately scaled and aligned
images \citep{2000ApJ...538L.113N}.  Thus, in order to make
predictions for the observable shapes of dark matter haloes from
simulations, we need to consider stacked projected shapes, rather than
the full 3-D triaxiality/sphericity distribution (as studied in
e.g. \citetalias{Bett07}).

The broad distribution of galaxy--halo (mis-)alignments
(Figs.~\ref{f:cosjdkmjgal_histo},
  \ref{f:cosjinnjgal_histo}, \ref{f:coscdkmjgal_histo}, 
  \ref{f:coscdkmcgal_histo} and \ref{f:cosadkmcgal_histo}) will have a significant impact on
estimates of halo shape from stacked 2-D mass distributions. To
quantify this, we first consider the population of relaxed,
well-resolved DMG haloes containing well-resolved
galaxies (i.e. $Q_\rmn{lim}=0.5$ and at least $300$ dark matter
particles in each halo, as well as at least $300$ star particles in
each galaxy).  Since we have access to the full particle
  coordinates, we can rotate each halo in 3-D such that the central
  galaxies' major--minor axes planes are all aligned.
We then compute the 2-D projected mass distribution matrix,
$\mat{M}$, of the halo, which has components:
\begin{equation}
  \mathcal{M}_{\alpha\beta} = \sum^{\Np}_{i=1} m_i r_{i,\alpha}r_{i,\beta},
\end{equation}
where the sum is over all particles in the halo (dark matter and
baryons), and $\alpha$, $\beta$ denote the matrix indices ($1$ or
$2$), such that $r_{i,1}$ is the halocentric distance of particle $i$
in the direction parallel to the galaxy's major axis and $r_{i,2}$ is
the distance parallel to the galaxy's minor axis. These matrices are
then normalised by the halo size, $\mat{M}^\prime =
\mat{M}/\left(\Mvir\Rvir^2\right)$.

Of course, this luxury -- rotating in 3-D, \emph{then} projecting --
is not available to observers, who are instead restricted to aligning
the projected galaxy images in 2-D.  Our method is not wholly
realistic therefore, but it does give an idealised `best-case' setup,
by allowing us to `observe' each galaxy--halo system with the galaxy
edge-on.  Furthermore, in practice, observers are likely
preferentially to select galaxies which appear more edge-on
\cite[e.g.][]{HYG2004,Parker2007}.

For each halo, we compute the eigenvectors ($\vv{a}$, $\vv{b}$) and
eigenvalues ($a>b$) of $\mat{M}^\prime$, giving the distribution of
projected halo shapes.  Finally, we take the mean of $\mat{M}^\prime$
over all the selected galaxies, and compute its eigenvectors and
eigenvalues.  These then describe the net shape distribution of the
selected haloes, in a 2-D projection, (perfectly) scaled and aligned
by their galaxy.

\begin{figure}
  \includegraphics[width=80mm]{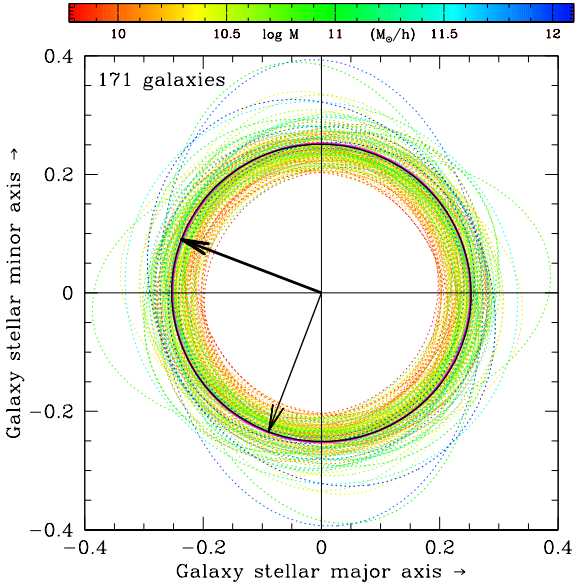} 
  \caption{The 2-D projected normalised mass distributions of DMG
    haloes (coloured dotted lines). Each halo has been aligned
    according to the orientation of its galaxy's stellar mass
    distribution.  The colouring is by the total halo mass, as
    indicated in the bar at the top.  The net result from stacking all
    haloes is the heavy black ellipse, with arrows marking the
    semimajor (heavy) and semiminor (lighter) axes.  Plotted beneath
    the stacked mass distribution ellipse is a circle (heavy magenta)
    as a visual aid.  The axes are labelled in the dimensionless units
    of the matrix $\mat{M}^\prime$ (see text).}
  \label{f:projmellipses}
\end{figure}

Fig. \ref{f:projmellipses} shows the ellipse defined by the
eigenvectors of the stacked system, along with those from each
individual halo.  Here we can see that although the projected mass
distribution of an individual halo is by no means necessarily
circular, the misalignment of the haloes with their galaxies results
in a stacked mass distribution that is almost exactly circular: the
axis ratio is $s=b/a=0.989$ (so the eccentricity is
$\epsilon=\sqrt{1-b^2/a^2} \simeq0.1$ and the ellipticity is $e =1-b/a
\simeq 0.01$). Given the uncertainties associated with measuring the
directions of the shape axes, this is indistinguishable from a
circular distribution.

\begin{figure}
  \includegraphics*[angle=-90,clip=true,trim=30mm 20mm 30mm 20mm,width=80mm]{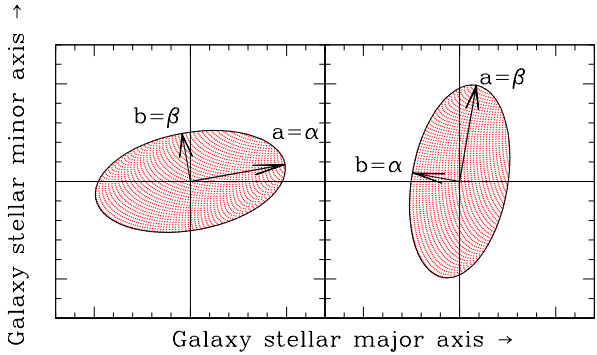} 
  \caption{Diagram illustrating the definition of the $\vv{\alpha}$
    and $\vv{\beta}$ projected halo shape axes, compared to the
    standard major and minor axes ($\vv{a}$ and $\vv{b}$; see
    Eqn.~\ref{e:alphabeta}).  The ellipse of a projected halo mass
    distribution is plotted onto axes that are aligned with its
    central galaxy.  We show two different possible orientations of
    the halo with respect to the galaxy: on the left, the halo and
    galaxy axes are nearly aligned, and on the right they are nearly
    perpendicular.  These two orientations show the two possible ways
    in which $\vv{\alpha}$ and $\vv{\beta}$ are assigned to $\vv{a}$
    and $\vv{b}$.}
  \label{f:littlediagram}
\end{figure}

\begin{figure}
  \includegraphics[width=80mm]{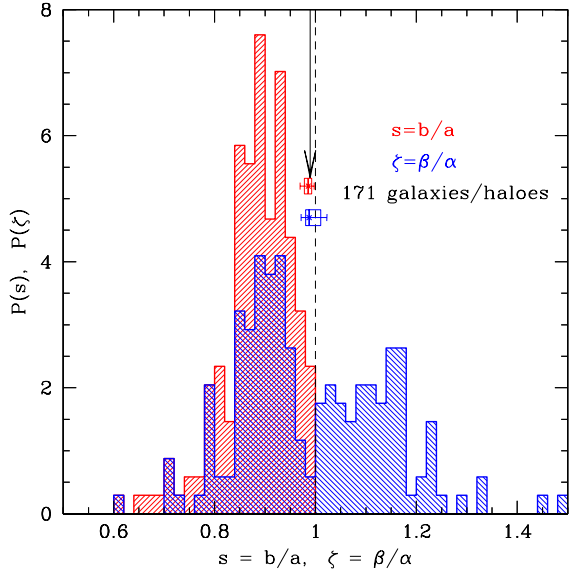} 
  \caption{Normalised histogram of the axis ratios $s$ (red) and
    $\zeta$ (blue) of the 2-D projected mass distributions of haloes
    shown in Fig.~\ref{f:projmellipses}.  Unity is marked with a
    dashed line.  The axis ratio of the stacked mass distribution is
    indicated by the arrow near $b/a\simeq 1$.  Below the arrow are
    two points ($s$ in red, $\zeta$ in blue, arbitrary heights) with
    percentile boxes and bars (as in
    Fig.~\ref{f:logr_logj_mbins_meds}), showing the spread of data
    from bootstrap resampling the projected halo distribution (see
    text).  This gives an estimate of the uncertainty on the stacked
    halo result. }
\label{f:projshapedistro}
\end{figure}

We can also label the projected halo axes according to which galaxy
axis they are closest to. The eigenvector lying closer to parallel to
the galactic major axis we label $\vv{\alpha}$, and that closer to the
galactic minor axis we label $\vv{\beta}$.  Thus, we have
\begin{eqnarray}
\vv{\alpha} = \left(\begin{array}{ll}\max\left(a_x,b_x\right) \\
\min\left(a_y,b_y\right) \end{array}\right), 
& \; & 
\vv{\beta} = \left(\begin{array}{ll}\min\left(a_x,b_x\right) \\
\max\left(a_y,b_y\right) \end{array}\right), 
\label{e:alphabeta}
\end{eqnarray}
where the $x$ and $y$ subscripts refer to the galactic major and minor
axes respectively (see Fig.~\ref{f:littlediagram} for a diagram).
Orthogonality means that the axis ratio $\zeta:=\beta/\alpha$ is equal
to either $s$ or $1/s$, depending on the galaxy--halo orientation.
The distributions of the individual and stacked axis ratios themselves
(both $s$ and $\zeta$) are shown in Fig. \ref{f:projshapedistro}.  Of
the $171$ haloes in the distribution, $77$ ($45$ per cent) have
$\zeta>1$, i.e. have their projected halo major axis more aligned with
the minor axis of their galaxy.  Even though the distributions peak
significantly away from $b/a=1$, the stacked result (marked by the
arrow) is essentially indistinguishable from unity.  We have estimated
the error on this stacked halo result, by bootstrap resampling the
projected halo shape data, and recomputing the stacked result.  The
two percentile boxes and bars in Fig. \ref{f:projshapedistro} show the
median and spread ($68$ per cent of the data within the boxes, $95$
per cent within the bars) of the data from $5000$ bootstrap
resamplings.

There have been several attempts to measure the shapes of dark matter
haloes from weak lensing data.  \cite{HYG2004} claimed to have found a
significant detection of halo ellipticity, excluding circular haloes
at the $99.5$ per cent confidence level, and yielding an average axis
ratio of $\approx 0.7$.  However, they lacked redshift data and this
limited the accuracy of their method. \cite{Parker2007} obtained a
similar result (an axis ratio of $\sim 0.7$ at a $2\sigma$ level) from
the CFHT Legacy Survey, again without redshift data.
\cite{Mandelbaum2006} used the large dataset of the SDSS (including
redshifts and morphologies), and performed a very thorough exploration
of possible systematic effects \cite[see also][]{Mandelbaum2005}.
Their results were far less conclusive than the previous studies, with
no definitive detection of halo ellipticity.  There was a hint that
spiral galaxies are aligned perpendicular to their haloes when
averaged over luminosities, and that ellipticals are increasingly well
aligned with increasing luminosity.  However, these results were not
statistically significant.  We have checked if our results depend on
galaxy morphology, as defined by the disc-to-total ratio
discussed earlier:  We are not able to detect any
statistically significant variation, although our sample sizes in this
case are small.

These studies highlight the practical difficulties in inferring halo
properties from weak lensing data.  The results we have presented here
show that the expected signal of non-circularity is extremely weak.
When assessing this conclusion, two caveats must be born in
mind. Firstly, the SO algorithm with which we identify haloes
inherently biases the halo mass distributions towards sphericity (see
\citetalias{Bett07}).  However, Figs.~\ref{f:projmellipses} and
\ref{f:projshapedistro} show that this is not the main cause of the
near spherical shape of the stacked halo. Rather, it is the
misalignment between galaxy and halo that smears out the significant
projected ellipticities of the haloes.

Secondly, we have only shown the projected halo ellipticities at the
virial radius.  Other studies have shown that CDM haloes become
increasingly aspherical towards the centre \citep{Hayashi2007}.
However, it is not clear that this is still the case for systems with
baryons: the limited studies so far of the shape profiles of CDM
haloes with baryons suggest that, regardless of mass, haloes could
become \emph{more} spherical towards the centre, depending on their
merger histories \citep{Kazantzidis2004, Bailin05,
  2006PhRvD..74l3522G, 2006ApJ...648..807B, 2009arXiv0902.2477A}.


\section{Conclusions}\label{s:conclusions}

In this paper we have investigated the angular momentum structure of
dark matter haloes in the \lcdm{} cosmology.  Our first and largest
sample includes $\sim 10^5$ well resolved haloes of galaxy, group and
cluster masses drawn from the Millennium Simulation and from a
smaller, higher resolution dark matter simulation (\hrsim{}). Our
second sample of $\sim 50$--$100$, also well resolved, galactic-size
haloes is drawn from two smaller simulations ran from the same initial
conditions, one with only dark matter and the other with baryons as
well (which can undergo cooling, star formation, feedback, etc.)

In the simulations without baryons, we have found that the median of
the cumulative specific angular momentum of dark matter as a function
of radius, $\jr$, scales approximately as $j\propto r$ (i.e. the
scaling that corresponds to circular motion in a mass distribution
with constant circular velocity). The amplitude of this trend scales
with halo mass, but even at a fixed mass, the scatter is large (over
an order of magnitude).  The dependence on halo mass is removed if
$\jr$ is scaled by $V_\rmn{vir}\Rvir$, i.e. 
$\jr/\sqrt{G\Mvir\Rvir}$. The angular velocity, $\omega(r)$, is
similarly independent of mass. These quantities, however, still
exhibit large scatter.  Thus, haloes do not rotate like solid bodies,
but have a rotation structure close to $\omega\propto r^{-1}$.  These
results apply over the $\sim 5$ orders of magnitude in halo mass
spanned by our pure dark matter simulations.

We investigated the coherence of the halo angular momentum by studying
how the orientation of the cumulative angular momentum vectors at
different radii deviate from that at an `inner' radius of $\approx
0.25\Rvir$, $\vjin$.  In the median, the total specific angular
momentum (i.e. of the mass within virial radius) is directed about
$25\degr$ away from $\vjin$.  Again there is large scatter:
Ninety-five per cent of the haloes have their total angular momenta
directed between $5\degr$ and $65\degr$ away from their $\vjin$.
There is a weak trend with mass, with the more massive haloes showing
a greater range of alignment angles (and greater misalignment in the
median). In the hierarchical model of structure formation, the more
massive haloes are more likely to have experienced a recent merger
event that could have altered their angular momentum structure.

We then investigated the effects that baryons have on the rotational
structure of haloes in the simulations described by
\cite{Libeskind07}, using the techniques
introduced by \cite{Takashi05}. Although the volume of this simulation
(diameter $\sim 12.5\lunit$) is too small to test if the galaxy luminosity
function is a good match to reality, at least the distribution of
morphological types is quite realistic: two thirds of the galaxies in
this simulation have $B$-band $D/T$ ratio greater than 0.5. In
parallel with this `DMG' simulation, we analysed its dark matter only
counterpart, `DMO'.

The main effect of the galaxy forming in the halo is to spin up the
inner parts, increasing the angular momentum of the region within
$0.1\Rvir$ by about $50$~per cent in the median. The increase becomes
smaller with radius and, at $\Rvir$, individual haloes simulated with
and without baryons are indistinguishable from each other. We have
shown that this increase is due, at least in part, to transfer of
angular momentum by gravitational tidal torques, from the baryons to
the dark matter. Overall the baryons do not conserve their own angular
momentum, particularly, as \cite{zavala2008} have shown, in
spheroid-dominated galaxies.

The process of galaxy formation also affects the coherence of the halo
angular momentum.  Although, again, there is a large amount of variation
amongst haloes, the median of the distribution indicates that,
overall, the formation of the galaxy helps to align the inner angular
momentum vector with that of the halo as a whole.  An analogous result
for the halo shape was obtained by \cite{Bailin05}.

Although the galaxy helps to align the inner halo with its outer
parts, the distribution of alignments is still very broad, ranging
from perfect alignment to misalignment of $\sim 120\degr$.
Furthermore, because of the large scatter and the relatively small
size of our galaxy sample, the distribution of alignments between the
inner and outer haloes (which is similar to that between the galaxy
and the halo in the DMG simulation) is statistically similar in our
simulations with and without baryons. We are also unable to
distinguish between the alignment distributions of bulge-dominated and
disc-dominated galaxies.

The minor axis of a galaxy is generally oriented parallel to the
direction of its angular momentum, even in slowly rotating,
bulge-dominated galaxies.  In simple models of galaxy formation, we
expect the angular momentum of the galaxy to be aligned with the shape
of its host halo. This expectation is the basis of the commonly used
method of stacking many galaxies together in order to amplify the weak
gravitational lensing signal produced by an individual galaxy, in
order to estimate, for example, halo ellipticity. In principle, the
shape of haloes provides a clean test of the \lcdm{} model since, as
many studies have shown, CDM haloes are generally strongly aspherical
\citep[e.g.][]{1988ApJ...327..507F,2002ApJ...574..538J,Bett07,Hayashi2007}.
In particular, such a test could distinguish \lcdm{} from modified
gravity theories, like TeVeS, in which the potential at large
distances from a galaxy should be spherical. Unfortunately, our
analysis suggests that this test will not work in practice.  The
galaxies are sufficiently misaligned with the shape of their haloes
that the stacking procedure will wash out any ellipticity signal in
the projected mass distribution. Further work with larger galaxy
samples is required to ascertain if this problem could be mitigated by
a careful sample selection (e.g. according to morphology or
luminosity).

In summary, a consistent picture for the spin and shape of \lcdm{}
haloes seems to be emerging from this and other studies. In the
absence of baryons, dark matter haloes are triaxial, with a preference
for prolateness, and become increasingly prolate towards their centres
\citep[e.g.][]{Bett07,Hayashi2007}.  They have very little coherent
rotation but they rotate with approximately constant rotational rather
than angular velocity (Figs.~\ref{f:logr_logj_mbins_meds} and
\ref{f:logr_angvel_medsall}).  A galaxy forming in the halo produces
several effects: the halo becomes more spherical overall, with a
tendency towards oblateness \citep{Kazantzidis2004, Bailin05,
  2006PhRvD..74l3522G, 2006ApJ...648..807B, 2009arXiv0902.2477A}; its
inner parts are spun up (Figs.~\ref{f:logr_angvel_medsall} and
\ref{f:logr_dmodmgmatchrat}), and become better aligned with the outer
parts, at $\Rvir$ (Figs.~\ref{f:logr_cosjrjin_dmodmg} and
\ref{f:logr_cosjdmodmg}). Thus, as they cool and collapse, the baryons
carry information about the outer halo orientation (at $\Rvir$) into
the inner halo (Figs.~\ref{f:logr_cosjgaljdmodmg},
\ref{f:cosjdkmjgal_histo} and \ref{f:cosjinnjgal_histo};
\citealt{Bailin05}, \citealt{2006PhRvD..74l3522G}). Nevertheless, the
alignment of the galaxy with the outer halo is weak and, furthermore,
there is very large halo-to-halo scatter in the size of the angular
momentum and in the alignment between the inner and outer halo. Half
of the galaxies have their minor axes inclined by more than $45\degr$
relative to that of their halo and about $10$ per cent of the galaxies
lie within $30\degr$ of the plane perpendicular to the halo major axis
(Figs.~\ref{f:coscdkmcgal_histo} and \ref{f:cosadkmcgal_histo}).

Although we have studied the largest set of simulated haloes and
galaxies available to date, two limitations of our galaxy sample
should be kept in mind. Firstly, even this sample is too small to
search for trends with morphology or formation history. Secondly, the
baryonic physics modelled in the simulations are extremely
uncertain. Even though our simulated galaxy population appears
realistic, at least in so far as the presence of different
morphological types is concerned, it includes only one of many
possible treatments of these fundamental, but complex, processes.


\section*{Acknowledgements}

PB acknowledges a PhD studentship from the Science \& Technology
Facilities Council (STFC), and the support of the Helmholtz Alliance
HA-101 `Physics at the Terascale'.  VRE is a Royal Society University
Research Fellow. CSF is a Royal Society Wolfson Research Merit Award
holder.  TO acknowledges support from the FIRST project based on
Grants-in-Aid for Specially Promoted Research by MEXT (16002003),
Grant-in-Aid for Scientific Research (S) by JSPS (20224002), and an
STFC Rolling Grant.  The simulations and analyses used in this paper
were carried out as part of the programme of the Virgo Consortium on
the Regatta supercomputer of the Computing Centre of the
Max-Planck-Society in Garching, and the Cosmology Machine
supercomputer at the Institute for Computational Cosmology, Durham.


\bibliographystyle{mn2e}
\bibliography{jr_dmgas}  


\appendix  
\section{Orientation resolution tests}\label{a:cosboot}
In Sections \ref{s:millorient} and \ref{s:dmodmgorient} we studied the
orientation of halo (or galaxy) angular momentum vectors.  These are
formed from a vector sum of the contributions from the objects'
constituent particles.  Because of the vector sum, the contribution
from many particles can be much smaller than that of a single
particle.  If the summed vector's magnitude is particularly small,
then the individual particles' vectors must have been largely in
opposite directions, and mostly cancelled out.  In that case, the
inclusion of very few particles with parallel vectors, or even a
single particle, can completely dominate the result.  Clearly, any
property that is dominated by discreteness effects is not reliable.
To ensure our results are robust, we have performed extensive Monte
Carlo tests using the halo catalogues from the \hrsim{}, DMO and DMG
simulations. (This problem has been tackled in the past by
\cite{2001ApJ...555..240B}, \cite{2004ApJ...616...27B}, and
\cite{2005ApJ...634...51A}.  We revisit it here to ensure that we can
retain as many haloes as possible, while rejecting those that are
unreliable.)

For each halo, we perform $5000$ bootstrap resamplings of both the
particles from the halo as a whole and also from just the halo inner
region, independently.  The cosine of the angle between the original
$\vjtot$ (or $\vjin$) and the bootstrap resampled $\vjtot$ (or
$\vjin$) is computed.  We take the median of the $5000$ samples for
each halo and plot it against the magnitude of the original vector,
rescaled in such a manner that any systematic trends are removed. For
this, we use the dimensionless quantity $\jsc(\le r) := \jr / j_0(r)$,
where $j_0(r)$ is the specific angular momentum of a (hypothetical)
test particle in a circular orbit at radius $r$. Since the circular
velocity $\vc(r) = \sqrt{G M(\le r) r^{-1}}$, we have:
\begin{equation}
  \jsc(\le r) = \frac{\jr}{\sqrt{G M(\le r) r} }
  \label{e:jsc}
\end{equation}
We have confirmed that this quantity does not vary systematically with
either mass or radius. (This $\jsc$ is similar to, but not the same
as, the scaled angular momentum shown in
Fig.~\ref{f:logr_logjmulti_dmodmg}.)

We wish to find a limiting value of $\jsc=:\jsc_\rmn{lim}$ such that
some given percentage of haloes with $\jsc\ge\jsc_\rmn{lim}$ have
their median bootstrapped vectors aligned to within a given angle.
Specifically, we require that $99.5$ per cent of the selected haloes
should have $\theta_\rmn{med} \le 15\degr$, where $\theta$ is the angle
between the actual halo $\vv{j}$ and the bootstrap resampled vector.

In practice, the limits obtained from this process, and the severity
of the cut on the resulting halo population, depend strongly on the
other selection criteria used.  We always restrict attention to
`virialised' haloes ($Q_\rmn{lim}=0.5$), but the minimum number of
particles in the given region (total or inner halo) has a strong
impact.  To balance the competing demands of a well resolved inner
region and spin orientations that are robust to discreteness effects,
we decided to move the `inner' radius outwards. While for the spin
magnitude analysis we adopted $r_\rmn{inner}=0.1\Rvir$, for the
orientation profiles we chose to move this two bins outwards to
$r_\rmn{inner} = 10^{-0.6}\Rvir \approx 0.25\Rvir$.  We also take
$1000$ as the minimum number of particles in the given region (total
or inner), rather than the $300$ we use for the analysis of the
angular momentum magnitude.  With these choices we were left with a
large enough sample of haloes to be statistically viable once the cuts
in $\Np$ and $\jsc$ had been applied.
 
\begin{figure}
  \includegraphics[width=80mm]{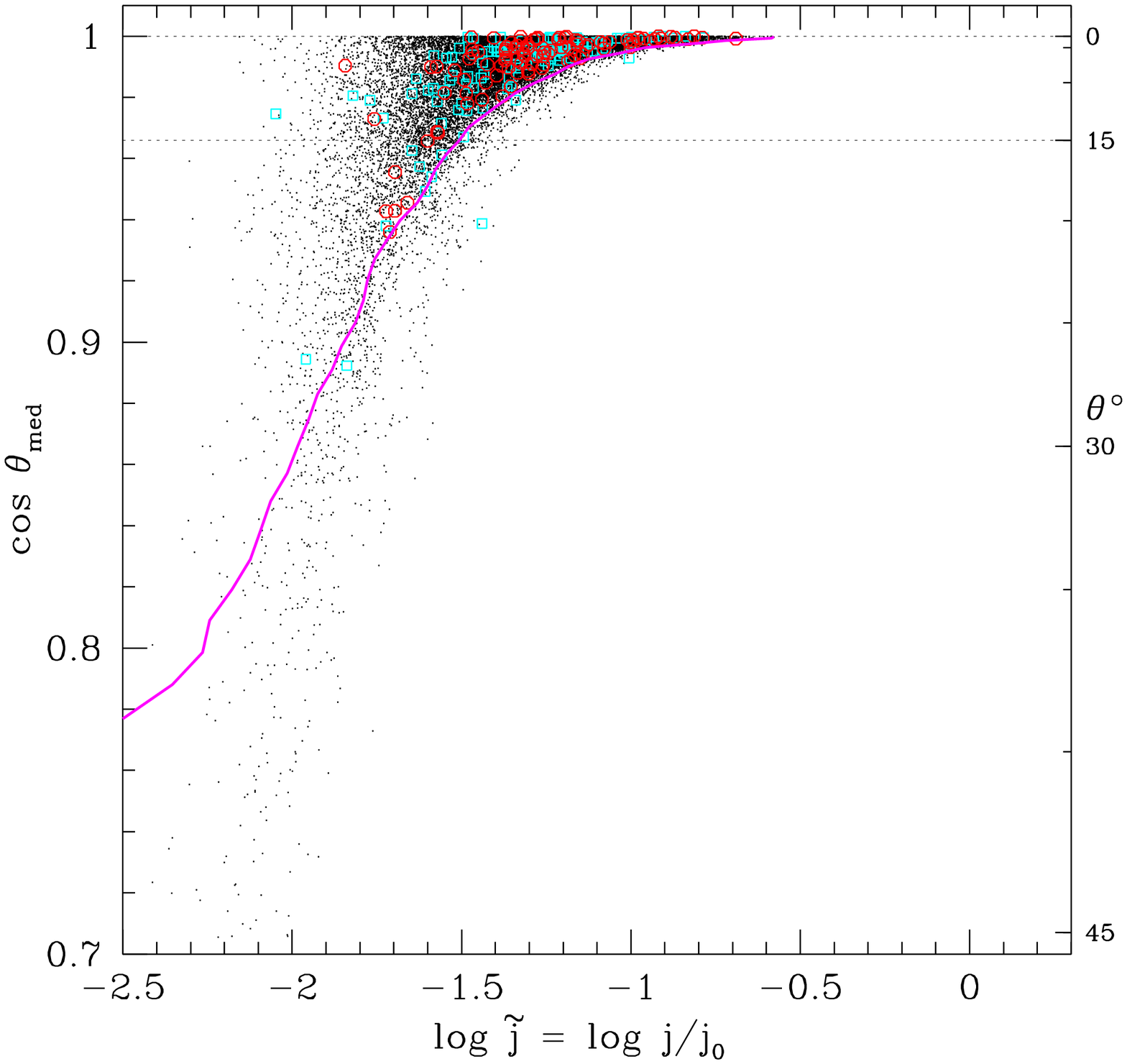} 
  \caption{Bootstrap resampling results for the angular momentum of
    the dark matter within $\Rvir$.  Each point is the median of the
    $5000$ angles between the $\vjtot$ of a halo and that of the
    bootstrap resamplings of that halo.  This angle is plotted against
    the halo's scaled angular momentum $\jsc$, so that the different
    simulations can be compared together.  The heavy magenta contour
    joins the series of lower limits on $\jsc$ such that $99.5$ per
    cent of the haloes with $\jsc\ge\jsc_\rmn{lim}$ are better aligned
    in the median than that angle.  The haloes have been selected to
    be `virialised' ($Q\le 0.5$) and well resolved (at least $1000$
    dark matter particles), and have been taken from the \hrsim{}
    (black dots), DMO (cyan squares) and DMG (red rings) simulations.
  }
  \label{f:cosboot_jtot}
\end{figure}

\begin{figure}
  \includegraphics[width=80mm]{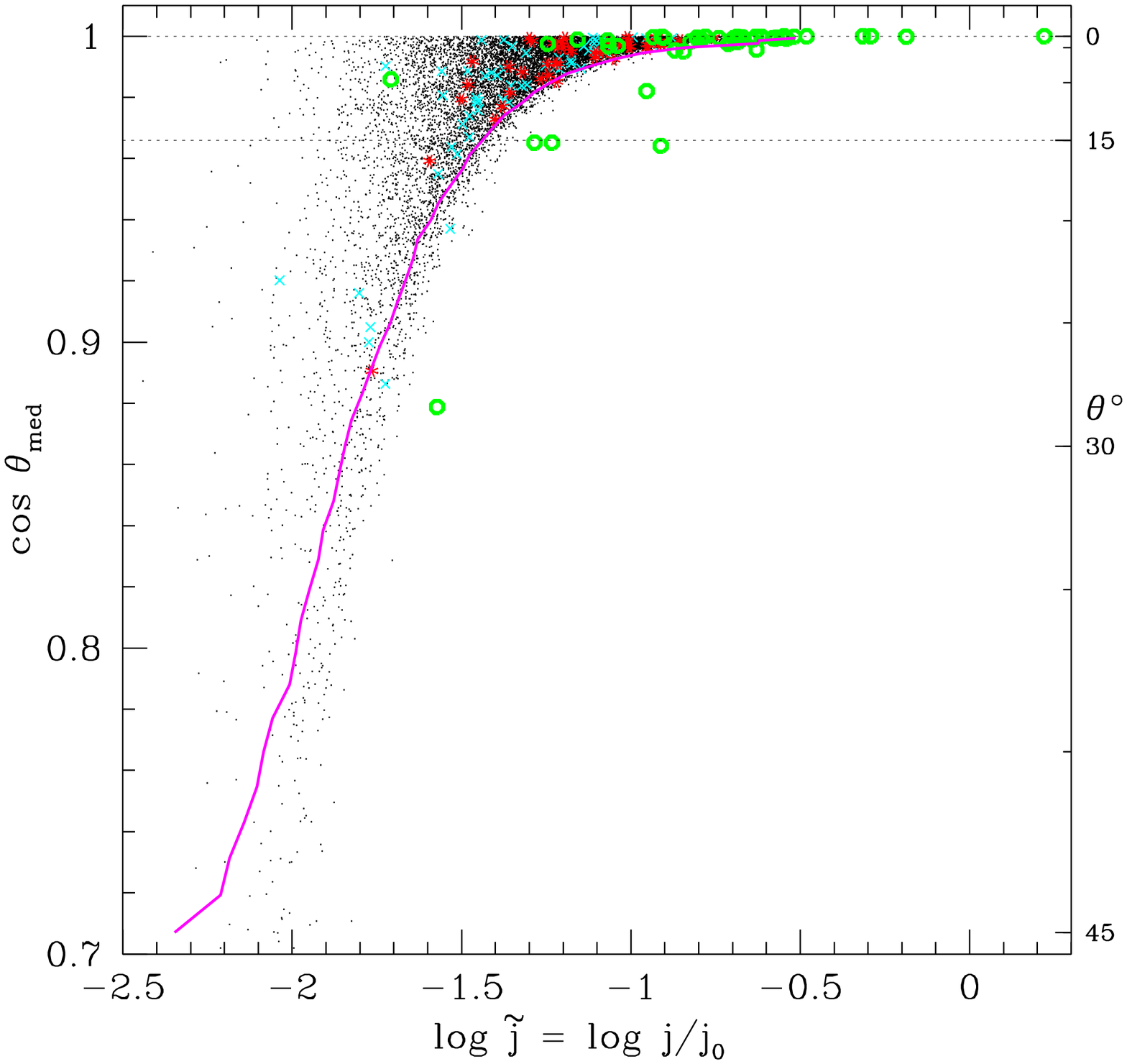} 
  \caption{As for Fig.~\ref{f:cosboot_jtot}, but for the angular
    momentum of the dark matter within $\approx 0.25\Rvir$; the $\Np$
    selection criterion therefore applies to this inner region.
    \hrsim{} haloes are marked as black dots, DMO haloes as blue
    crosses, and DMG haloes as red asterisks.  Also included are the
    angular momenta of the stellar components of the central galaxies
    from the DMG haloes (green rings).  To be included, the DMG
    haloes/galaxies need also to satisfy $\Np\ge 1000$ for the star
    particles.  Inclusion of the galaxies does not affect the
    resulting limit on $\jsc$.}
  \label{f:cosboot_jinn}
\end{figure}

Fig. \ref{f:cosboot_jtot} shows the bootstrap results for the angular
momentum of dark matter within $\Rvir$.  The median of the
bootstrapped angles from each halo are plotted. There is a clear
increase in scatter for haloes with low $\jsc$ (that is, low $j$
compared to that of a typical particle).  The same trend is present
for haloes from the \hrsim{}, DMO and DMG simulations.  Our
requirement that most haloes should have a median bootstrap angle
within $15\degr$ translates into a selection criterion of
approximately $\log_{10}\jsc\ge -1.5$.  Fig. \ref{f:cosboot_jinn}
shows the results of applying this method to the inner regions of the
haloes only.  Although the scatter behaves slightly differently, the
value of $\jsc_\rmn{lim}$ is very similar.  It is worth noting that,
as long as they are selected with the same criteria, the inclusion of
the DMO and DMG haloes does not affect the value of $\jsc_\rmn{lim}$
significantly.  Including the stellar components of galaxies does not
affect the results greatly either, since the galaxies generally have
higher angular momentum anyway.

Thus, the halo selection criteria we adopt for our analysis of angular
momentum orientation are:
\begin{eqnarray*}
  Q & \le & 0.5 \\
  \Np & \ge & 1000 \: \mbox{(inner or total)}\\
  \log_{10}\jsc & \ge & \left\{\begin{array}{lr}
      -1.44 & \mbox{(total)} \\
      -1.51 & \mbox{(inner)} 
    \end{array}\right.
\end{eqnarray*}
We do not select according to the $\jsc$ of the galaxies.

\begin{figure}
  \includegraphics[width=80mm]{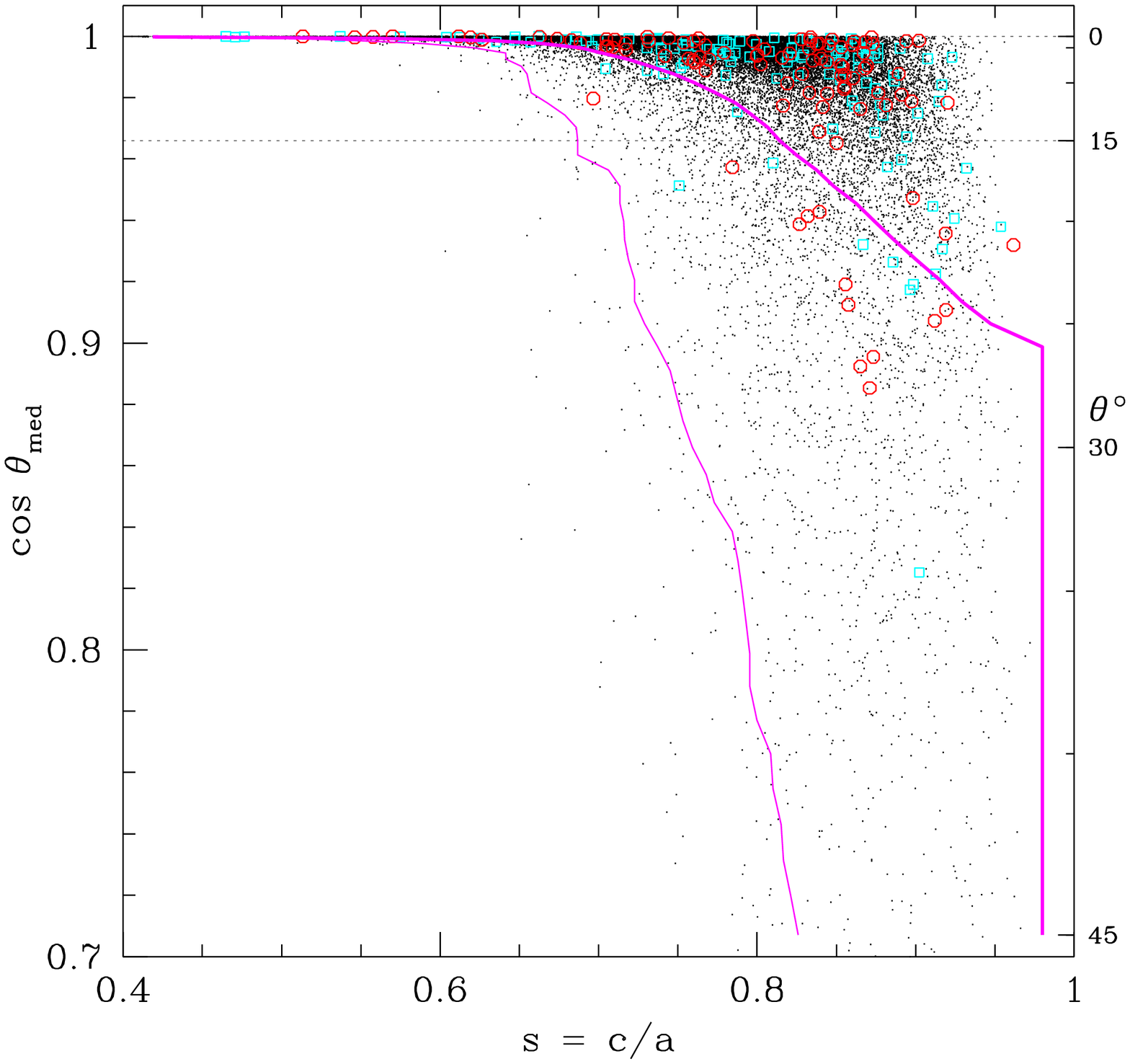} 
  \caption{The angle between the measured and bootstrapped halo major
  axes plotted against the minor-to-major axis ratio, $s$, using
  similar principles to those in Fig~\ref{f:cosboot_jtot}. Due to the
  greater scatter in the angles in this case, we plot two
  contours. One (thin) is such that $99.5$ per cent of the haloes that
  have a larger value of $s$ are better aligned than the given angle;
  the other (thick) corresponds to the $95.0$ per cent level.  The
  larger scatter results in much more restrictive cuts. }
\label{f:cosboot_smax}
\end{figure}

A similar problem arises when considering the orientation of the dark
matter mass distribution, i.e. the halo shape given by the mutually
orthogonal vectors $\vv{a}$, $\vv{b}$ and $\vv{c}$.  Regarding the
axis ratio $s=c/a$ as a measure of how close the halo is to spherical,
it is clear that if $s$ is very close to unity, then the three axes
become degenerate and the halo orientation is undetermined.  Thus,
just as for angular momentum, we use bootstrap resampling to find a
limiting value of $s$ ($s_\rmn{max}$) such that an acceptable fraction
of the haloes with $s\leq s_\rmn{max}$ have sufficiently reliable
directions (Fig.~\ref{f:cosboot_smax}).  We find that the shape axes
scatter more under bootstrapping than the angular momentum vectors. If
we want to retain a significantly large sample we have to be less
demanding on the accuracy of their orientations.  We reduce the
percentile limit to be such that $95$ per cent of haloes with $s\leq
s_\rmn{max}$ have their median bootstrap angle within $15\degr$.
This yields a limit of $s_\rmn{max} = 0.81$ which we apply as an
additional selection criterion in Figs.~\ref{f:cosctotjin_histo}, 
\ref{f:coscdkmjgal_histo}, and~\ref{f:coscdkmcgal_histo}.

\label{lastpage}
\end{document}